\documentclass[12pt]{iopart}

\usepackage{amssymb}
\usepackage{mathrsfs} 					
\usepackage[french,english]{babel} 
\usepackage[dvips]{graphicx}			

\begin{document}

\title[]{{\it Ab initio} calculation of the 66 low lying electronic states of HeH$^+$: adiabatic and diabatic representations}

\author{J\'er\^ome Loreau$^1$, Jacques Li\'evin$^1$, Patrick Palmeri$^2$, Pascal Quinet$^{2,3}$ and Nathalie Vaeck$^1$}

\address{$^1$ Laboratoire de Chimie Quantique et Photophysique, Facult\'e des Sciences, Universit\'e Libre de Bruxelles,
50, av. F. Roosevelt, CP160/09, 1050 Bruxelles, Belgium.}
\address{$^2$ Service d'Astrophysique et de Spectroscopie, Universit\'e de Mons, 20 Place du Parc, B-7000 Mons, Belgium.}
\address{$^3$ 3 IPNAS, Universit\'e de Li\`ege, Bat.15, Sart Tilman, B-4000 Li\`ege, Belgium.}
\ead{nvaeck@ulb.ac.be}

\begin{abstract}
We present an {\it ab initio} study of the HeH$^+$ molecule. Using the quantum chemistry package MOLPRO and a large adapted basis set, we have calculated the adiabatic potential energy curves of the first 20 $^1 \Sigma^+$, 19 $^3\Sigma^+$, 12 $^1\Pi$, 9 $^3\Pi$, 4 $^1\Delta$ and 2 $^3\Delta$ electronic states of the ion in CASSCF and CI approaches. The results are compared with previous works. The radial and rotational non-adiabatic coupling matrix elements as well as the dipole moments are also calculated. The asymptotic behaviour of the potential energy curves and of the various couplings between the states is also studied.
Using the radial couplings, the diabatic representation is defined and we present an example of our diabatization procedure on the $^1\Sigma^+$ states.

\end{abstract}
\submitto{\JPB}



\maketitle

\section{Introduction}

HeH$^+$, or hydrohelium cation, is thought to be the first molecular species to appear in the Universe, its formation being due to radiative association between H$^{+}$ and He (Roberge \& Dalgarno 1982). In addition, the high fractional abundance of HeH$^+$ should allow its detection in stars formed from primordial material such as the recently discovered very metal-poor stars HE1327-2326 and HE0107-5240 (Frebel \etal 2005) or in He-rich environment as in the white dwarfs SDSS J133739+000142 and LHS 3250 (Harris \etal 2004). The inclusion of HeH$^+$ in the existing atmospherical models of those objects could have serious implications. It should also be present in the planetary nebula NGC7027 but has eluded observation (Moorhead \etal 1988). In fact, up to now, none of the several attempts to extraterrestrial observation of HeH$^+$ have been conclusive (Engel \etal 2005). 
Considering that the excited states of HeH$^+$ are too shallow or unstable to support a visible or UV spectrum, those assessments have risen up a number of studies to obtain theoretically and experimentally the most accurate rotational spectrum of HeH$^+$ in the ground state, culminating with the recent work of Stanke \etal (Stanke \etal 2006). In addition to extremely accurate knowledge of the spectroscopic properties of the hydrohelium cation, the various mechanisms leading to its formation or decay must be investigated to obtain a correct estimation of the population of the levels. In this context, the first experimental data for the photodissociation cross section in the far UV has been obtained recently using the free electron laser FLASH at Hambourg (Pedersen \etal 2007), showing important disagreement with the previous theoretical works and motivating new calculations (Sodoga \etal 2009, Dimitriu \& Saenz 2009). 

Despite the fact that its astrophysical observation is still questionable, HeH$^+$ is clearly present in helium-hydrogen laboratory plasmas. Indeed, since its first observation in mass spectrometry of discharges in mixtures of helium and hydrogen in 1925 (Hogness \& Lunn, 1925), HeH$^+$ has been found to be one of the major components in other He/H plasma sources such as high voltage glow discharges, synchrotron devices, inductively coupled plasma generators, capacitively coupled RF discharges, and magnetically confined plasmas, the last one playing of course a very special role in today's development of thermonuclear fusion. 
Helium emission lines have been proposed recently as a diagnostic tool for divertor regions of the tokamak. However, to model the intensity of these emission lines, a knowledge of the cross sections of the charge transfer processes which populate the emitting levels of helium up to $n=4$ is essential. At low or very low collisional energy, the theoretical description of charge transfer requires a molecular approach and the calculation of the excited states of the HeH$^+$ quasi molecule (Rosmej \etal 2006).

From a theoretical point of view, the hydrohelium cation is the simplest closed-shell heteronuclear molecule and therefore a considerable amount of work has been dedicated to high precision calculations of its ground state, including accurate description of relativistic and non-adiabatic effects (Stanke \etal 2008). In a lesser measure, the first excited states have also been used to assess the efficiency of different {\it ab initio} methods to describe states that are not the lowest of their symmetry (Richings \& Karadakov 2007) or to understand and remedy the failure of time-dependent density functional theory to provide accurate charge transfer excitation energies (Giesbertz \etal 2008). The most complete study of the excited states of HeH$^+$ has been performed by Green \etal in a series of four articles (Green \etal 1974a, 1974b, 1976, 1978). States up to $n=3$ (where $n$ is the highest principal quantum number in the dissociation configuration of hydrogen or helium) have been calculated using a CI (Configuration Interaction) method with a combination of Slater-type orbitals and ellipsoidal orbitals. Despite the fact that this exhaustive study includes the calculation of dipole matrix elements and radial non-adiabatic couplings, its accuracy has never been assessed and the use of these results directly in quantum molecular dynamics programs is problematic due to the lack of data for medium or large internuclear distances.

While the first excited states of the neutral HeH molecule emanate from the excitation of the hydrogen atom alone, the first part of the electronic spectrum of HeH$^+$ results from a mixing between states arising from single excitations of neutral helium or hydrogen. The higher part of the spectrum is built upon single excitations of the He$^+$ cation as well as double excitations of neutral helium and of the ground state of the H$^-$ anion. 

The purpose of this article is to reexamine the first part of the electronic spectrum of HeH$^+$ molecule and to extend its description up to $n=4$ with high-level {\it ab initio}  quantum chemistry methods in order to provide adequate material required for both spectroscopy and dynamical studies such as charge transfer processes in excited states (Loreau \etal 2010) or photodissociation in the far UV domain (Sodoga \etal 2009). In addition, the diabatic representation of the potential energy curves is investigated. The results of the present work are compared to the corresponding data in the literature, when available. 

All the data described in this paper are accessible on demand to the corresponding author of this article.

\section{One electron basis set and asymptotic atomic energy levels}\label{atomic_energies}

One problem encountered in this work is the construction of a reliable gaussian basis set allowing the description of the formation of the HeH$^+$ molecular ion from the first Rydberg states of the H and He atoms up to $n=4$. As mentioned above, the first part of the electronic spectrum corresponds indeed asymptotically to the excitation of both hydrogen and neutral helium. Therefore, our basis set consists for each atom of the aug-cc-pV5Z basis set (Dunning, 1989; Woon and Dunning, 1994) augmented by [$3s,3p,2d,1f$] Gaussian type orbitals optimized to reproduce the spectroscopic orbitals of the He and H excited states. A different atomic basis has been developed up to $n=4$ for He depending on whether the molecular state under consideration is a singlet or a triplet state. Those additional sets of orbitals have been obtained by fitting Slater type orbitals from calculations performed for each atomic state using the AUTOSTRUCTURE package (Eissner \etal 1974, Badnell 1986, 1997). In total, a [8s,7p,5d,3f,1g] basis set as been used for both atoms. The additional orbitals and their contraction coefficients are given in the appendix. 

For all values of $\ell$, this basis set reproduces the exact non-relativistic atomic levels of hydrogen up to $n=3$ within 15 cm$^{-1}$. The electronegativity of H$^-$ deviates from the experimental value from 37 cm$^{-1}$  in a full CI level of approximation.
Different gaussian basis sets have already been proposed in the literature mainly for the calculation of the ground state of HeH$^+$ (Jurek \etal 1995 and references therein) or the ground and excited states of the neutral HeH molecule which correspond asymptotically to excitations of H up to $n=3$ (van Hemert and Peyerimhoff 1991). Using this last atomic basis set, the levels for the hydrogen atom are reproduced within 20 cm$^{-1}$ for the $s$ and $p$ states and within 70 cm$^{-1}$ for the $d$ state. 

In the non-relativistic approximation, the $1s\ nl$ ($n=1-3$) levels of helium are described at a full CI level by our basis set within 115 cm$^{-1}$ for $s$ states, 60 cm$^{-1}$ for $p$ states and 30 cm$^{-1}$ for $d$ states. 

For both atoms, the $s$ states for $n=4$ are more difficult to reproduce, mainly due to the lack of upper states and the large number of lower states, but are still in a reasonable agreement (145 cm$^{-1}$ at most) with the exact values. The other $n=4$ states are reproduced within 42 cm$^{-1}$ for both atoms.

In conclusion, although this basis set is rather small, it is adapted to the HeH$^+$ system and will allow a correct description of the potential energy curves as well as the determination of the non-adiabatic couplings which, in our approach, require CASSCF (Complete Active Space Self-Consistent Field) calculations.


\begin{table}
\caption{\label{TableSingSig} $^1\Sigma^+$ states included in the calculations and their dissociation product. In this table, we use the notation H($nl$) or He($1snl\ ^1L$) to denote the electronic wavefunctions of the corresponding electronic states. 
Due to the He$^+$ charge, there is a Stark effect on the hydrogen levels (see text). The mixing coefficients have been calculated by diagonalizing the perturbation matrix due to the electric field.}
\begin{indented}
\item[]
\begin{minipage}{\textwidth}
\begin{tabular}{@{}cccc}
\br
	& $m$	& Energy (hartree)	& Dissociative atomic wavefunctions \footnote{It is understood that H($nl$) is accompanied by He$^+(1s)$ and that He($1s nl\ ^1L$) \\ is accompanied by H$^+$.} \\
	&		& at $R=70$ a.u.	& \\
\mr
$n=1$	& 1	& -2.90324307 & {He}($1s^2$) \\
		& 2	& -2.49995502 & H($1s$) \vspace{1mm} \\ 
$n=2$	& 3	& -2.14589424 & He($1s2s\ ^1S$) \\
		& 4	& -2.12556499 & $\frac{1}{\sqrt{2}}$ H($2s$) + $\frac{1}{\sqrt{2}}$ H($2p$)\\
		& 5	& -2.12433765 & $\frac{1}{\sqrt{2}}$H($2s$) - $\frac{1}{\sqrt{2}}$H($2p$) \\
		& 6	& -2.12374055 & He($1s2p\ ^1P^{o}$) \vspace{1mm} \\
$n=3$	& 7	& -2.06157066 & He($1s3s\ ^1S$)  \\
		& 8	& -2.05758300 & $\frac{1}{\sqrt{3}}$H($3s$) - $\frac{1}{\sqrt{2}}$H($3p$) + $																\frac{1}{\sqrt{6}}$H($3d$) \\ 
		& 9	& -2.05632793 & He($1s3d\ ^1D$) \\
		& 10 & -2.05537040 & $\frac{1}{\sqrt{3}}$H($3s$) - $\sqrt{\frac{2}{3}}$H($3d$) \\
		& 11 & -2.05379172 & He($1s3p\ ^1P^{o}$) \\
		& 12 & -2.05411889 & $\frac{1}{\sqrt{3}}$H($3s$) + $\frac{1}{\sqrt{2}}$H($3p$) + $																\frac{1}{\sqrt{6}}$H($3d$) \vspace{1mm} \\
$n=4$	& 13	& -2.03701879 & He($1s4s\ ^1S$) \\
		& 14 & -2.03502013 & $\frac{1}{2}$H($4s$) - $\frac{3}{2 \sqrt{5}}$H($4p$) + $\frac{1}												{2}$H($4d$) - $\frac{1}{2 \sqrt{5}}$H($4f$) \\
		& 15 & -2.03266813 & He($1s4f\ ^1F^{o}$)  \\
		& 16 & -2.03194805 & $\frac{1}{2}$H($4s$) - $\frac{1}{2 \sqrt{5}}$H($4p$) - $\frac{1}													{2}$H($4d$) + $\frac{3}{2 \sqrt{5}}$H($4f$) \\
		& 17 & -2.03027998 & He($1s4d\ ^1D$)  \\
		& 18 & -2.02961868 & $\frac{1}{2}$H($4s$) + $\frac{1}{2 \sqrt{5}}$H($4p$) - $\frac{1}													{2}$H($4d$) - $\frac{3}{2 \sqrt{5}}$H($4f$) \\
		& 19 & -2.02851159 & He($1s4p\ ^1P^{o}$)  \\
		& 20 & -2.02802390 & $\frac{1}{2}$H($4s$) + $\frac{3}{2 \sqrt{5}}$H($4p$) + $\frac{1}{2}$H($4d$) + $\frac{1}{2 \sqrt{5}}$H($4f$) \\
\br
\end{tabular}
\end{minipage}
\end{indented}
\end{table}

\begin{table}
\caption{\label{TableTripSig}$^3\Sigma^+$ states states included in the calculations and their dissociation product. In this table, we use the notation H($nl$) or He($1snl\ ^3L$) to denote the electronic wavefunctions of the corresponding electronic states. 
Due to the He$^+$ charge, there is a Stark effect on the hydrogen levels (see text). The mixing coefficients have been calculated by diagonalizing the perturbation matrix due to the electric field.}
\begin{indented}
\item[]
\begin{minipage}{\textwidth}
\begin{tabular}{@{}cccc}
\br
	& $m$	& Energy (hartree)	& Dissociative atomic wavefunctions \footnote{It is understood that H($nl$) is accompanied by He$^+(1s)$ and that He($1s nl\ ^3L$) \\ is accompanied by H$^+$.} \\
	&		& at $R=70$ a.u.	& \\
\mr
$n=1$	& 1	& -2.49996040 & H($1s$) \\
$n=2$	& 2	& -2.17513428 & He($1s2s\ ^3S$) \\
		& 3	& -2.13288467 & He($1s2p\ ^3P^{o}$) \\
		& 4	& -2.12557046 & $\frac{1}{\sqrt{2}}$H($2s$) + $\frac{1}{\sqrt{2}}$H($2p$)) \\
		& 5	& -2.12434310 & $\frac{1}{\sqrt{2}}$H($2s$) - $\frac{1}{\sqrt{2}}$H($2p$) \vspace{1mm} \\
$n=3$	& 6	& -2.06880105 & He($1s3s\ ^3S$) \\
		& 7 	& -2.05841457 & He($1s3p\ ^3P^{o}$) \\
		& 8	& -2.05758856 & $\frac{1}{\sqrt{3}}$H($3s$) - $\frac{1}{\sqrt{2}}$H($3p$) + $																\frac{1}{\sqrt{6}}$H($3d$) \\
		& 9	& -2.05537599 & $\frac{1}{\sqrt{3}}$H($3s$) - $\sqrt{\frac{2}{3}}$H($3d$) \\
		& 10	& -2.05520640 & He($1s3d\ ^3D$) \\
		& 11 & -2.05379735 & $\frac{1}{\sqrt{3}}$H($3s$) + $\frac{1}{\sqrt{2}}$H($3p$) + $																\frac{1}{\sqrt{6}}$H($3d$) \vspace{1mm} \\
$n=4$	& 12	& -2.03814067 & He($1s4s\ ^3S$) \\
		& 13 & -2.03562121 & He($1s4p\ ^3P^{o}$) \\
		& 14 & -2.03292454 & $\frac{1}{2}$H($4s$) - $\frac{3}{2 \sqrt{5}}$H($4p$) + $\frac{1}												{2}$H($4d$) - $\frac{1}{2 \sqrt{5}}$H($4f$) \\
		& 15 & -2.03195557& $\frac{1}{2}$H($4s$) - $\frac{1}{2 \sqrt{5}}$H($4p$) - $\frac{1}													{2}$H($4d$) + $\frac{3}{2 \sqrt{5}}$H($4f$) \\
		& 16 & -2.02962842 & He($1s4d\ ^3D$) \\
		& 17 & -2.02955929 & $\frac{1}{2}$H($4s$) + $\frac{1}{2 \sqrt{5}}$H($4p$) - $\frac{1}													{2}$H($4d$) - $\frac{3}{2 \sqrt{5}}$H($4f$) \\
		& 18 & -2.02803385 & $\frac{3}{2 \sqrt{5}}$H($4p$) + $\frac{1}{2}$H($4d$) + $\frac{1}{2 \sqrt{5}}$H($4f$) \\
		& 19 & -2.01655785 & He($1s4f\ ^3F^{o}$) \\
\br
\end{tabular}
\end{minipage}
\end{indented}
\end{table}

\begin{table}
\caption{\label{TableSingPi} $^1\Pi$ states included in the calculations and their dissociation product. In this table, we use the notation H($nl$) or He($1snl\ ^1L$) to denote the electronic wavefunctions of the corresponding electronic states. 
Due to the He$^+$ charge, there is a Stark effect on the hydrogen levels (see text). The mixing coefficients have been calculated by diagonalizing the perturbation matrix due to the electric field.}
\begin{indented}
\item[]
\begin{minipage}{\textwidth}
\begin{tabular}{@{}cccc}
\br
	& $m$	& Energy (hartree)	& Dissociative atomic wavefunctions \footnote{It is understood that H($nl$) is accompanied by He$^+(1s)$ and that He($1s nl\ ^1L$) \\ is accompanied by H$^+$.} \\
	&		& at $R=70$ a.u.	& \\
\mr
$n=2$	& 1	& -2.12491660 & H($2p$) \\
		& 2	& -2.12368473 & He($1s2p\ ^1P^{o}$) \vspace{1mm} \\ 
$n=3$	& 3	& -2.05639837 & $\frac{1}{\sqrt{2}}$H($3p$) + $\frac{1}{\sqrt{2}}$H($3d$) \\
		& 4	& -2.05616425 & He($1s3d\ ^1D$) \\
		& 5	& -2.05456333 & $\frac{1}{\sqrt{2}}$H($3p$) - $\frac{1}{\sqrt{2}}$H($3d$) \\
		& 6	& -2.05419837 & He($1s3p\ ^1P^{o}$) \vspace{1mm} \\
$n=4$	& 7	& -2.03379403 & $\frac{1}{\sqrt{3}}$H($4p$) - $\frac{1}{\sqrt{2}}$H($4d$) + $																\frac{1}{\sqrt{6}}$H($4f$) \\ 
		& 8	& -2.03345358 & He($1s4f\ ^1F^{o}$) \\
		& 9	& -2.03081819 & $\frac{1}{\sqrt{3}}$H($4p$) - $\sqrt{\frac{2}{3}}$H($4f$) \\
		& 10 & -2.03065750 & He($1s4d\ ^1D$) \\
		& 11 & -2.02892558 & $\frac{1}{\sqrt{3}}$H($4p$) + $\frac{1}{\sqrt{2}}$H($4d$) + $																\frac{1}{\sqrt{6}}$H($4f$)  \\
		& 12 & -2.02877573 & He($1s4p ^1P^{o}$) \\
\br
\end{tabular}
\end{minipage}
\end{indented}
\end{table}

\begin{table}
\caption{\label{TableTripPi} $^3\Pi$ states included in the calculations and their dissociation product. In this table, we use the notation H($nl$) or He($1snl\ ^3L$) to denote the electronic wavefunctions of the corresponding electronic states. 
Due to the He$^+$ charge, there is a Stark effect on the hydrogen levels (see text). The mixing coefficients have been calculated by diagonalizing the perturbation matrix due to the electric field.}
\begin{indented}
\item[]
\begin{minipage}{\textwidth}
\begin{tabular}{@{}cccc}
\br
	& $m$	& Energy (hartree)	& Dissociative atomic wavefunctions \footnote{It is understood that H($nl$) is accompanied by He$^+(1s)$ and that He($1s nl\ ^3L$) \\ is accompanied by H$^+$.} \\
	&		& at $R=70$ a.u.	& \\
\mr
$n=2$	& 1	& -2.13282525 & He($1s2p\ ^1P^{o}$) \\
		& 2	& -2.12491845 & H($2p$) \vspace{1mm} \\ 
$n=3$	& 3	& -2.05814410 & He($1s3p\ ^1P^{o}$) \\
		& 4	& -2.05640023 & $\frac{1}{\sqrt{2}}$H($3p$) + $\frac{1}{\sqrt{2}}$H($3d$) \\
		& 5	& -2.05526105 & He($1s3d\ ^1D$) \\
		& 6	& -2.05456518 & $\frac{1}{\sqrt{2}}$H($3p$) - $\frac{1}{\sqrt{2}}$H($3d$) \vspace{1mm} \\
$n=4$	& 7	& -2.03371307 & He($1s4p ^1P^{o}$) \\ 
		& 8	& -2.03323584 & $\frac{1}{\sqrt{3}}$H($4p$) - $\frac{1}{\sqrt{2}}$H($4d$) + $																\frac{1}{\sqrt{6}}$H($4f$) \\
		& 9	& -2.03081642 & $\frac{1}{\sqrt{3}}$H($4p$) - $\sqrt{\frac{2}{3}}$H($4f$) \\
\br
\end{tabular}
\end{minipage}
\end{indented}
\end{table}

\begin{table}
\caption{\label{TableSingDelta}$^1\Delta$ states included in the calculations and their dissociation product. In this table, we use the notation H($nl$) or He($1snl\ ^1L$) to denote the electronic wavefunctions of the corresponding electronic states. 
Due to the He$^+$ charge, there is a Stark effect on the hydrogen levels (see text). The mixing coefficients have been calculated by diagonalizing the perturbation matrix due to the electric field.}
\begin{indented}
\item[]
\begin{minipage}{\textwidth}
\begin{tabular}{@{}cccc}
\br
	& $m$	& Energy (hartree)	& Dissociative atomic wavefunctions \footnote{It is understood that H($nl$) is accompanied by He$^+(1s)$ and that He($1s nl\ ^1L$) \\ is accompanied by H$^+$.} \\
	&		& at $R=70$ a.u.	& \\
\mr
$n=3$	& 1	& -2.05540649 & H($3d$) \\
		& 2	& -2.05537712 & He($1s3d\ ^1D$) \\
$n=4$	& 3	& -2.03206818 & He($1s4f\ ^1F^{o}$) \\
		& 4	& -2.03204221 & $\frac{1}{\sqrt{2}}$H($4d$) + $\frac{1}{\sqrt{2}}$H($4f$)) \\
\br
\end{tabular}
\end{minipage}
\end{indented}
\end{table}

\begin{table}
\caption{\label{TableTripDelta}$^3\Delta$ states included in the calculations and their dissociation product. In this table, we use the notation H($nl$) or He($1snl\ ^3L$) to denote the electronic wavefunctions of the corresponding electronic states. 
Due to the He$^+$ charge, there is a Stark effect on the hydrogen levels (see text). The mixing coefficients have been calculated by diagonalizing the perturbation matrix due to the electric field.}
\begin{indented}
\item[]
\begin{minipage}{\textwidth}
\begin{tabular}{@{}cccc}
\br
	& $m$	& Energy (hartree)	& Dissociative atomic wavefunctions \footnote{It is understood that H($nl$) is accompanied by He$^+(1s)$ and that He($1s nl\ ^3L$) \\ is accompanied by H$^+$.} \\
	&		& at $R=70$ a.u.	& \\
\mr
$n=3$	& 1	& -2.05543591 & He($1s3d\ ^1D$) + H$^+$ \\
		& 2	& -2.05540833 & He$^+(1s)$ + H($3d$) \\
\br
\end{tabular}
\end{minipage}
\end{indented}
\end{table}

\section{Potential energy curves}\label{PEC}

The Born-Oppenheimer adiabatic potential energy curves (PEC) for the lowest molecular states corresponding asymptotically to excitation in the $n = 1-4$ atomic shells have been calculated as a function of the internuclear distance $R$ using the MOLPRO molecular structure package (Werner \etal 2006). This includes 20 $^1\Sigma^+$, 19 $^3\Sigma^+$, 12 $^1\Pi$, 9 $^3\Pi$, 4 $^1\Delta$ and 2 $^3\Delta$ states, which constitutes a total of 66 electronic states. All these states, as well as their energy at $R=70$ a.u. and their dissociation products, are presented in Tables \ref{TableSingSig} to \ref{TableTripDelta}. To obtain the PEC, we performed a state-averaged CASSCF (Werner and Knowles 1985, Knowles and Werner 1985) using the active spaces listed in table \ref{AS}, followed by a full CI.

As we will not consider any spin-dependent interaction, the singlet and triplet states can be calculated separately.

\begin{table}
\caption{Active spaces used in state-averaged CASSCF calculations.}\label{AS}
\begin{indented}
\item[]
\begin{tabular}{@{}cccccc}
\br
$n$ value & $^1\Sigma^+$ & $^3\Sigma^+$ & $^1\Pi$ & $^3\Pi$ & $^{1,3}\Delta$ \\
\mr
1 \ & ($6\sigma,2\pi$) ($X^1\Sigma^+$) & ($9\sigma,4\pi,1\delta$) &  &   &  \\
   \ & ($9\sigma,4\pi,1\delta$) ($A^1\Sigma^+$) & &  &   &  \\
2 \ & ($9\sigma,4\pi,1\delta$) & ($9\sigma,4\pi,1\delta$) & ($9\sigma,4\pi,1\delta$) & ($9\sigma,4\pi,1\delta$) & \\
3 \ & ($12\sigma,6\pi,4\delta$)	& ($12\sigma,8\pi,2\delta$) & ($4\sigma,8\pi,6\delta$) & ($4\sigma,10\pi,2\delta$) & ($6\sigma,4\pi,4\delta$) \\
4 \ & ($28\sigma,2\delta$) & ($28\sigma$) & ($2\sigma,13\pi,2\delta$) &  & ($6\sigma,4\pi,4\delta$) \\
\br
\end{tabular}
\end{indented}
\end{table}

\subsection{Adiabatic $n=1$ potential energy curves}

\begin{figure}[h]
\centering
\includegraphics[angle=-90,width=15cm]{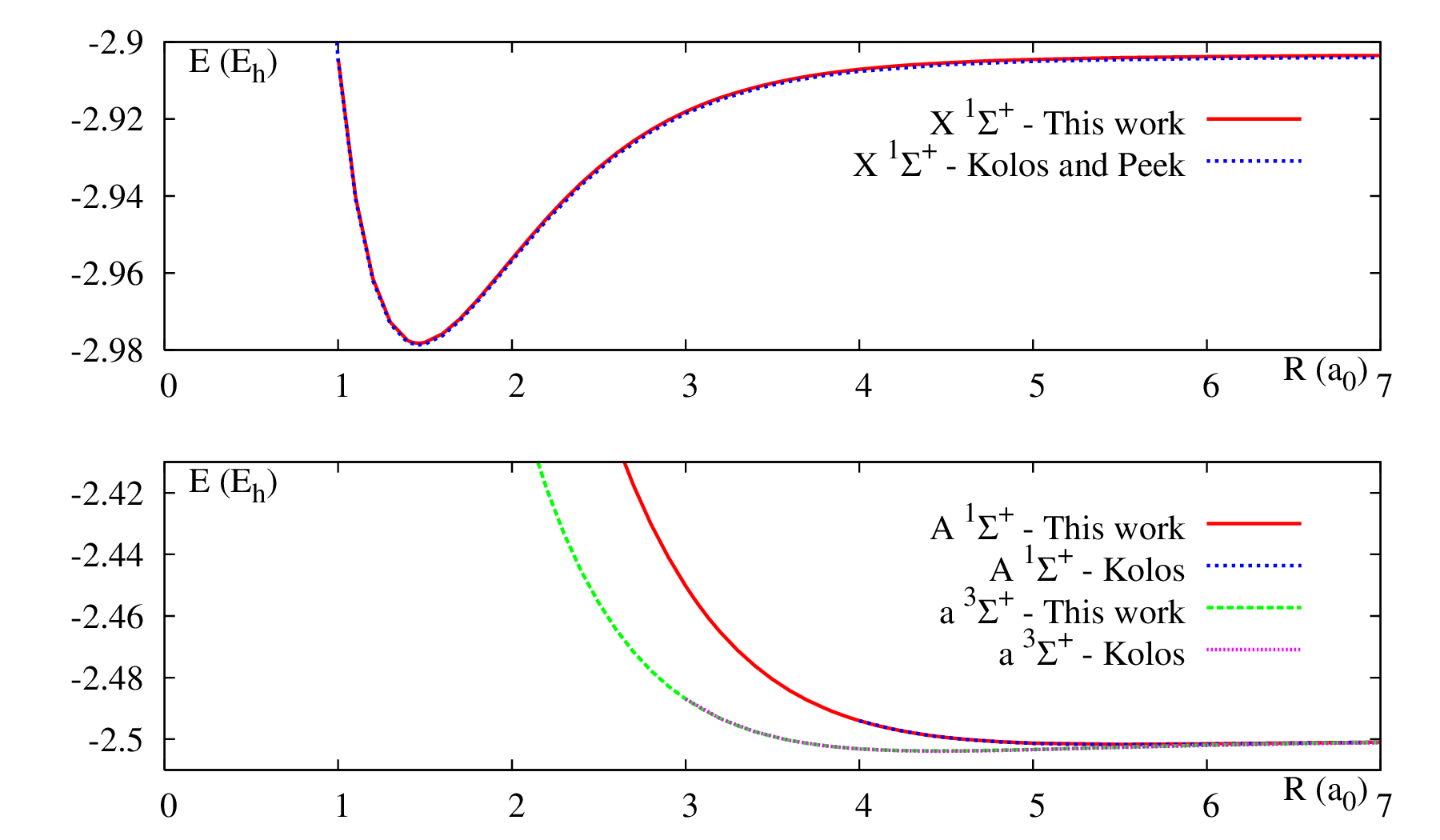}
\caption{Adiabatic potential energy curves (in hartrees) of the $n=1$ states. Comparison with the work of Kolos (1976) and Kolos \& Peek (1976).}
\label{n=1states}
\end{figure}

The $n=1$ states consist of two $^1\Sigma^+$ and one $^3\Sigma^+$ states and are shown in figure \ref{n=1states}. All three are bound states that support a vibrational structure.

Our calculations reproduce correctly the equilibrium distance of 1.463 a.u. of the $X\ ^1\Sigma^+$ state.
The dissociation energy of the ground state calculated by Kolos and Peek (1976) is $D_{e}$ = 16 455.64 cm$^{-1}$ and a more accurate value of 16 456.51 cm$^{-1}$, which include diagonal Born-Oppenheimer corrections, was given by Bishop and Cheung (1979). From an experimental point of view, a numerical procedure particularly successful when data are fragmentary has been employed by Coxon and Hajigeorgiou (1999) to inverse the spectroscopic line positions of the ground state potential of HeH$^+$. A remarkable agreement has been obtained with the theoretical values of Bishop and Cheung (1979).

The value obtained in this work for the dissociation energy is of 16 464 cm$^{-1}$. The energy depth in our calculation is therefore less than 9 cm$^{-1}$ too shallow and is to be compared to the result of van Hemert and Peyerimhoff (1990) which is 298 cm$^{-1}$ too large. Following these authors, this discrepancy is indicative of a basis set deficiency. The Basis Set Superposition Effect (BSSE) has been evaluated by the counterpoise method and has been found to be negligible.

In addition, using a $B$-spline basis set method, we have resolved the vibrational nuclear equation for $^4$HeH$^+$ and obtained 12 vibrational bound states, as was found in the recent paper of Stanke \etal (2006).

The $A \;^1\Sigma^+$ and the $a \; ^3\Sigma^+$ states have been studied by Kolos (1976) in the adiabatic approximation. They both present weakly attractive potential curves with a minimum at large internuclear distances, $R_e = 5.53$ a.u. and $4.47$ a.u. for the singlet and the triplet (respectively), to be compared with our values of 5.53 a.u. and 4.45 a.u. We have determined the dissociation energies $D_{e}=849.71$ cm$^{-1}$ for the $A$ state and $D_{e}=379.70$ cm$^{-1}$ for the $a$ state.
A study of the vibrational structure of the $a\ ^3\Sigma^+$ state of $^4$HeH$^+$ has been performed by Chibisov \etal (1996) using the potential energy curve of Michels (1966) extended by an analytical expression in the asymptotic region. They found that this potential supports 5 bound vibrational levels, while our resolution of the vibrational motion produced 6 bound levels. However, the binding energy of the last level is less than 1 cm$^{-1}$. The largest difference between our energy values and those of Chibisov \etal is of 2 cm$^{-1}$.
It is also important to note that the non-adiabatic couplings between the $^3\Sigma^+$ states have been neglected in both calculations and could modify significantly the energy of the levels, as it has been shown already for the vibrational structure of the ground state (Stanke \etal 2006).
Finally, in our calculations, we found that 4 bound vibrational levels are supported by the $A\ ^1\Sigma^+$ state of $^4$HeH$^+$.

\subsection{Adiabatic potential energy curves for the first Rydberg $^{1,3}\Sigma^+$ states}

\begin{figure}[h!]
\includegraphics[angle=-90,width=15cm]{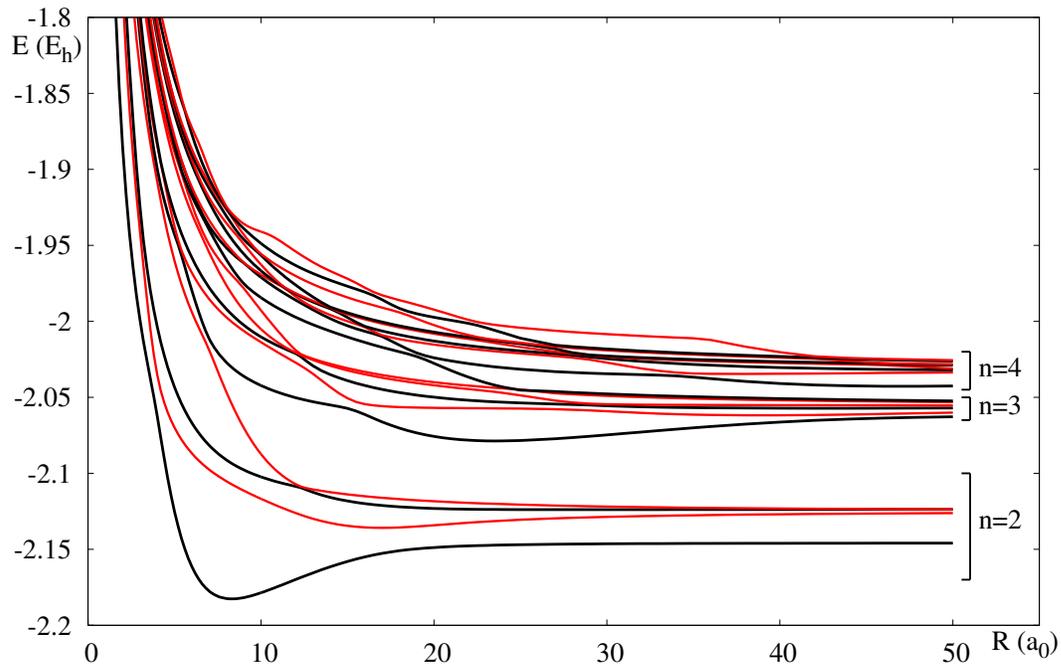}
\caption{Adiabatic potential energy curves for the $n=2-4$ $^1\Sigma^+$ states. In black, states dissociating into He($1snl\ ^1L$) + H$^+$. In red, states dissociating into He$^+(1s)$ + H($nl$).}
\label{pot1}
\end{figure}

\begin{figure}[h]
\includegraphics[angle=-90,width=15cm]{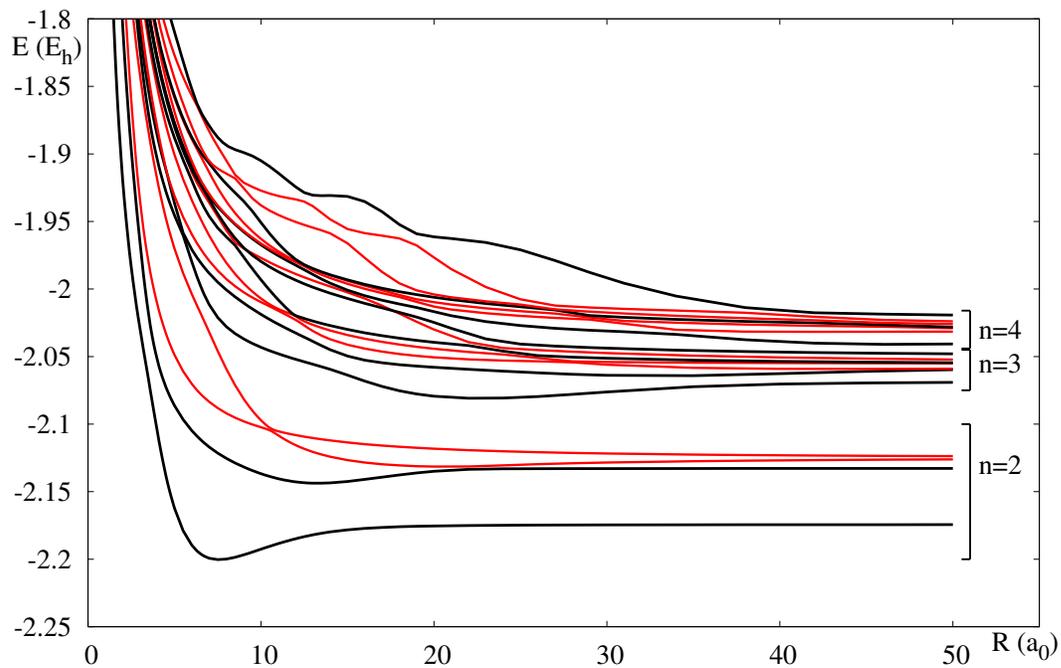}
\caption{Adiabatic potential energy curves for the $n=2-4$ $^3\Sigma^+$ states. In black, states dissociating into He($1snl\ ^3L$) + H$^+$. In red, states dissociating into He$^+(1s)$ + H($nl$).}
\label{pot2}
\end{figure}

The $n=2-4$ states of HeH$^+$ can be divided in two groups. The first category dissociated asymptotically into H$^+$ and neutral He in an excited $1s\ n\ell \; ^{1,3} L$ state, while the second one dissociated into He$^+$ in its $1s$ ground state and an excited $n\ell$ state of H. Both categories of states alternated along the electronic spectrum resulting in a large number of avoided crossings leading to charge exchange dynamics.
For the $\Sigma$ symmetry, the total number of states up to $n=4$ is 20 for the singlets and 19 for the triplets. They are shown in figures \ref{pot1} and \ref{pot2}, respectively.
In these figures, the states of the first category are drawn in black while the states of the second category are in red.

The $n=2$ manifolds have a very similar behaviour in both singlet and triplet spin symmetries. However, for the singlet states an avoided crossing between the two highest states occurs at an internuclear distance of 50 a.u. 
This crossing is understandable once looking at the asymptotical behaviour of those two $^1 \Sigma^+$ states, which is governed by the Stark effect. Indeed, at large internuclear distances, the system is composed of a neutral atom, perturbed by an ion. The atomic dissociative states are He($1s2p \; ^{1}P^o$) + H$^+$ and He$^+ (1s$) + H($2s$), respectively for the highest and the lowest state.

If we restrict the description to quadratic effects in the field, the helium state behaves asymptotically in the presence of the H$^+$ charge as
\begin{equation}
E(R) = E^0_{\scriptstyle{He} (1s2p \; ^1P^o)} -  \frac{\alpha_{He (1s2p \; ^1P^o)}}{2R^4}
\end{equation}
and the $2s$ state of H, under the electric field produced by He$^+$, as
\begin{equation}
E(R) = E^0_{He^+ (1s) + H(2s)} + \frac{3}{R^2} - \frac{\alpha_{H(2s)}}{2R^4}
\end{equation}
where $E^0$ represent the atomic energies. In addition, the $\vert 2s \rangle$ state of hydrogen becomes $\frac{1}{\sqrt{2}}\vert 2s\rangle - \frac{1}{\sqrt{2}}\vert 2p\rangle$.

The first-order Stark effect produces a term proportional to $1/R^2$, which vanishes unless the atomic state is degenerate with a state of opposite parity (Goldman and Cassar 2005).  The term proportional to $1/R^4$ is due to the second-order Stark effect, and the constant $\alpha$ is the dipole polarizability. The polarizabilities for the helium states can be found in Yan (2000, 2002) while the polarizabilities for hydrogen are obtained analytically (Radzig and Smirnov 1985). 
The result is that while the helium state presents an almost flat asymptotic curve, the hydrogen state decreases to its atomic value, and a crossing occur in the analytical model at $R=50$ a.u., almost exactly as in the {\it ab initio} calculation. 

\begin{figure}[h!]
\includegraphics[angle=-90,width=15cm]{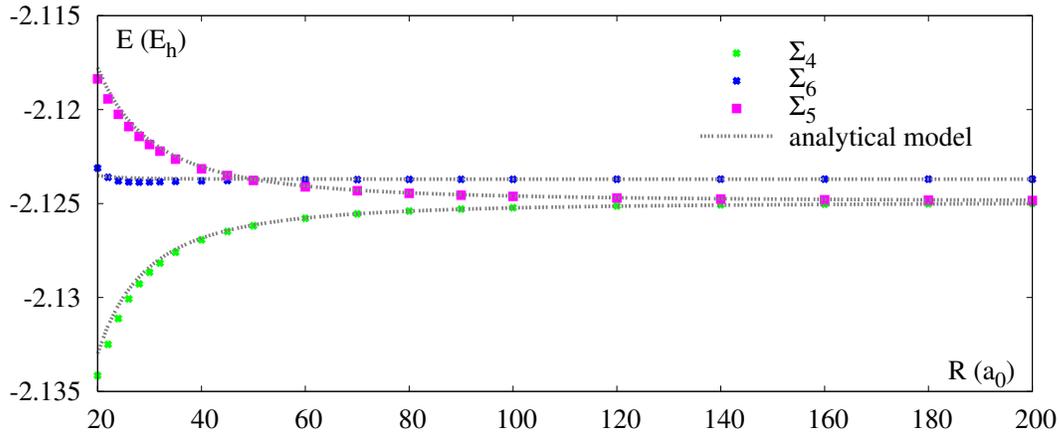}
\caption{Asymptotic behaviour of the three highest $n=2$ $^1\Sigma^+$ states. The numbers in subscript refer to the value of $m$ as defined in table \ref{TableSingSig}.}
\label{starkn=2}
\end{figure}

\begin{figure}[h!]
\includegraphics[angle=-90,width=15cm]{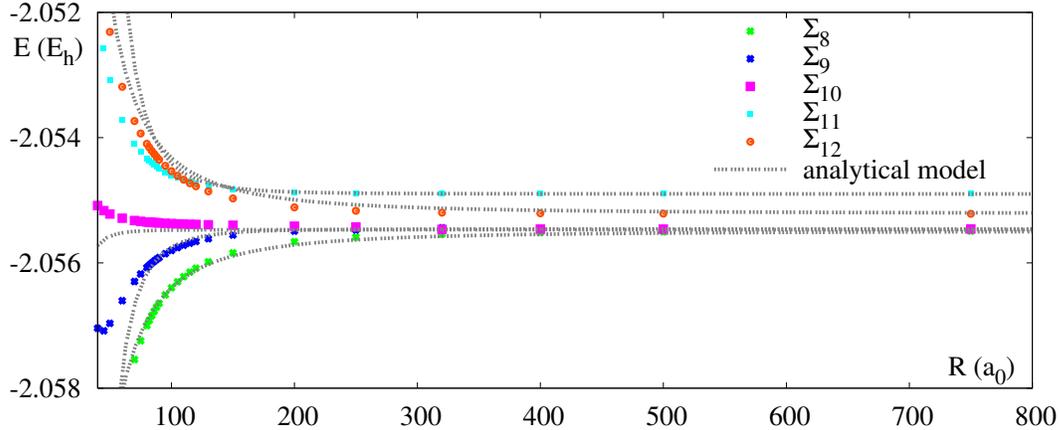}
\caption{Asymptotic behaviour of the five highest $n=3$ $^1\Sigma^+$ states.}
\label{starkn=3}
\end{figure}

From figure \ref{starkn=2}, we see that the asymptotic $n=2$ states are correctly described using the Stark effect up to order 2. To a lesser extent, it is also the case of the $n=3$ singlet states (see figure \ref{starkn=3}): for example, the analytical model reproduces the crossing which occurs at 90 a.u. between the eleventh and twelfth states dissociating respectively into He$^+(1s)$ + H($3s$) and He$(1s3p \; ^1P)$ + H$^+$ (see table \ref{TableSingSig}).

In the Born-Oppenheimer approximation, in which the quantum chemical calculations are performed, these crossings are avoided. 
However, the large amplitude and the narrowness of the non-adiabatic couplings at those points indicates that a full diagonal diabatic representation at the crossing is perfectly justified.


It is clear from the figures \ref{pot1} and \ref{pot2} that the number of avoided crossings increases strongly with $n$ and that their positions are shifted to larger internuclear distances.
Those avoided crossings mask partially the intrinsic oscillatory behaviour of the higher states that has been observed in previous calculations (Boutalib and Gadea, 1992). This behaviour has been related to the nodal structure of the Rydberg orbitals which may start to be important in the $n=4$ manifold and can be clearly seen in the last two $^3\Sigma^+$ states.
The second observation is that the states belonging to different manifolds are very close in energy, especially at internuclear distances smaller than 10 a.u. In addition, well-defined avoided crossings couple the last state of a manifold to the first state of the next manifold at large internuclear distances (7 a.u. between $n=2$ and $n=3$ and 18 a.u. between $n=3$ and $n=4$ for the singlet symmetry).

Finally, due to the presence of the $n=5$ states, very close in energy and not adequately described in our calculations, the representation of the highest $n=4$ states is probably less accurate than for the other members of the Rydberg series.

The comparison can be made with the previous work of Green \etal (Green \etal 1974a, 1974b, 1976, 1978) for the $n=2$ singlet and triplet states in the range $R=1-5$ a.u. and with the data of Kl¨\"uner \etal (1999) for the 3 first singlet states in the range $R=1-10$ a.u. The different sets of results are very similar for both spin symmetry, the present calculations being systematically more stable in energy.

We also note that Kl¨\"uner \etal (1999) have associated, at short internuclear distances, the last state of the $n=2$ manifold in their calculations as dissociating into He($1s2p\ ^1P^o$) + H$^+$ while due to the crossing involving this state at 50 a.u., it actually corresponds to He$^+$($1s$) + H($n=2$) (see table \ref{TableSingSig}). This is very important since these authors are interested in the electron transfer mechanism and they have eliminated the last $n=2$ state from their dynamical calculations.

For the $n=3$ manifold, only Green \etal provide a full set of results for both singlet and triplet states. The differences with our data are more important than for the $n=2$ states in both spin multiplicities, especially for the highest states which undergo avoided crossings with the lowest $n=4$ states.

Tables \ref{TabNacmeSig} and \ref{TabNacmePi} give the location of all the major avoided crossing points for singlet and triplet states which have been determined by an analysis of the calculated radial non-adiabatic couplings (see section \ref{radcoupl}).

\subsection{Adiabatic potential energy curves for the first Rydberg $^{1,3}\Pi$ and $^{1,3}\Delta$ states}

\begin{figure}[h!]
\includegraphics[angle=-90,width=15cm]{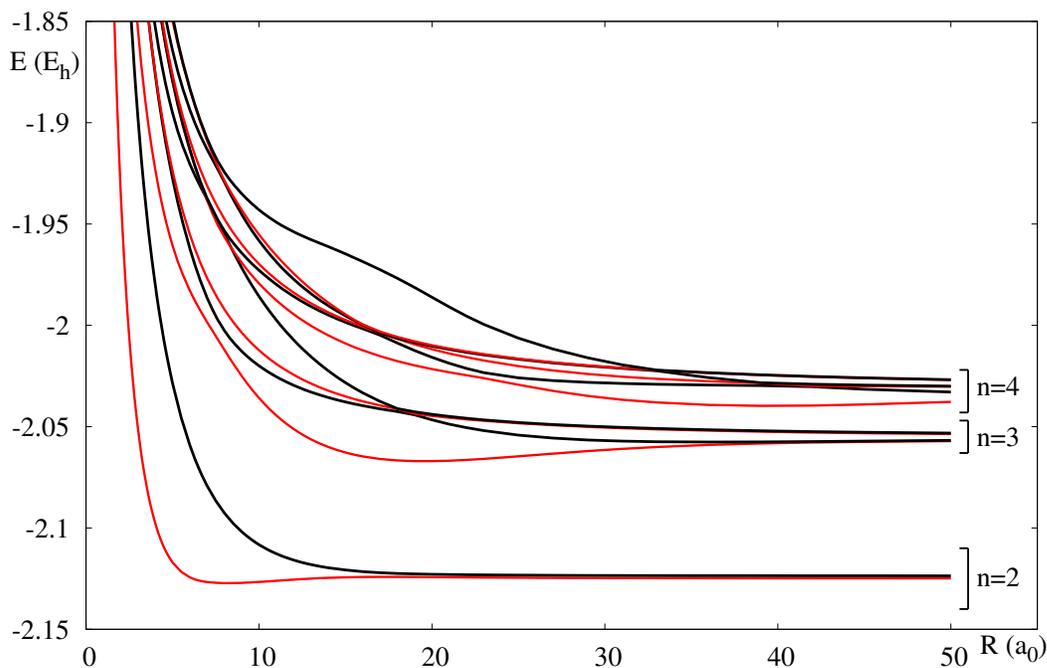}
\caption{Adiabatic PEC of the $n=2-4$ $^1\Pi$ states. In black, states dissociating into He($1snl\ ^1L$) + H$^+$. In red, states dissociating into He$^+(1s)$ + H($nl$).}
\label{singpi}
\end{figure}

\begin{figure}[h!]
\includegraphics[angle=-90,width=15cm]{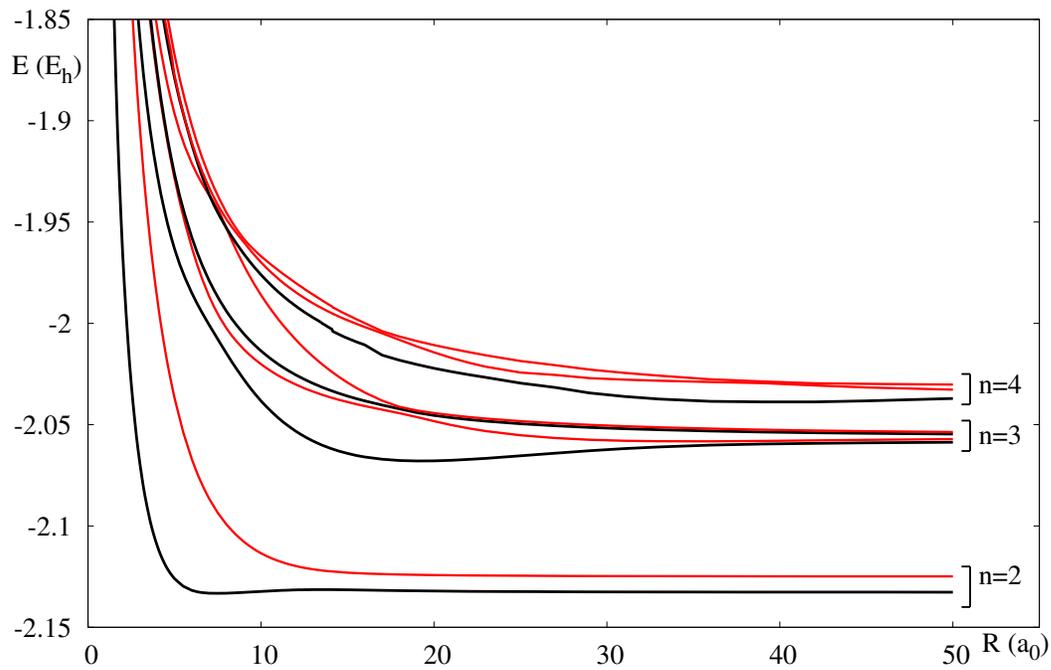}
\caption{Adiabatic PEC of the $n=2-4$ $^3\Pi$ states. In black, states dissociating into He($1snl\ ^3L$) + H$^+$. In red, states dissociating into He$^+(1s)$ + H($nl$).}
\label{trippi}
\end{figure}

The potential energy curves of the $^1 \Pi$ and $^3 \Pi$ states are presented in figures \ref{singpi} and \ref{trippi}, respectively. 
The PEC have again a very similar behaviour for both spin multiplicities. Indeed, the first and the third state seem to present a shallow well, and we also see that an avoided crossing between the last three $n=3$ states occurs at $R=18$ a.u. Finally, the last $n=3$ state and the first $n=4$ state interact strongly at about 8 a.u.

Once again, the qualitative comparison with the results of Green \etal for $n=2-3$ is good but there are some differences. The avoided crossing mentioned above occurs at $R=20$ a.u. rather than at $R=18$ a.u. in our work, and we also see that the energy separation at the avoided crossing between the third and fourth state (internuclear distance of about 8 a.u.) is larger in our calculation.

\begin{figure}[h!]
\includegraphics[angle=-90,width=15cm]{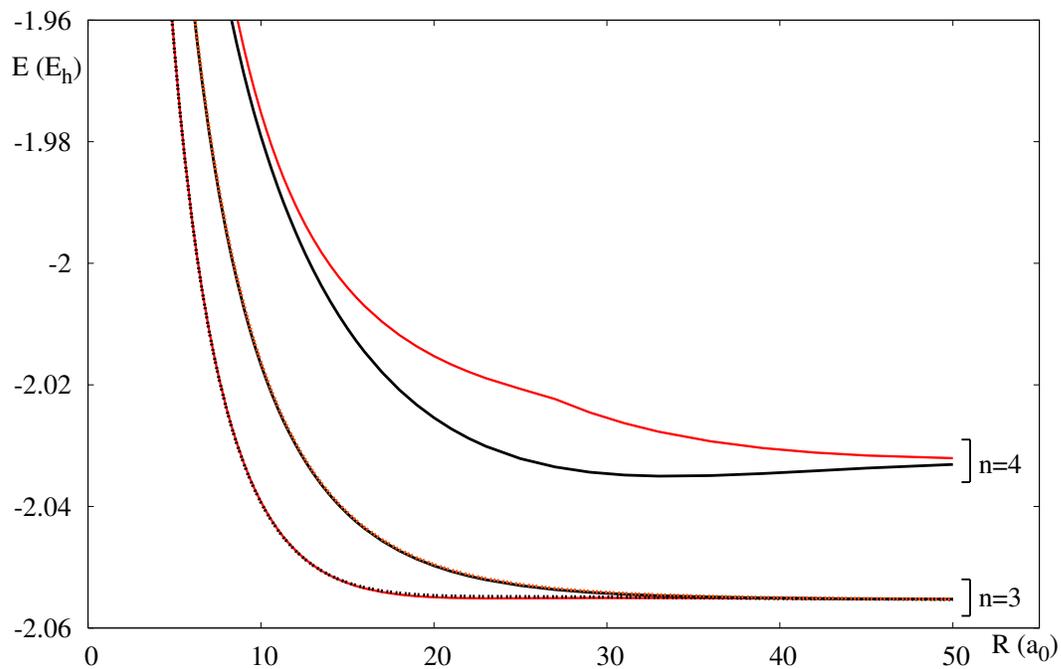}
\caption{Adiabatic PEC of the $n=3-4$ $^1\Delta$ and of the $n=3$ $^3\Delta$ states (full and dashed lines, respectively). In black, states dissociating into He($1snl\ ^{1,3}L$) + H$^+$. In red, states dissociating into He$^+(1s)$ + H($nl$).}
\label{delta}
\end{figure}

Finally, the PEC for the $\Delta$ states are presented in figure \ref{delta}. They seem to be almost independent of the spin multiplicity. In the work of Green \etal, the two $n=4$ PEC present an avoided crossing, which is not the case in this work.

It should be noted that the calculation of the $\Delta$ states is more difficult than for the other symmetries. This is due to the fact that MOLPRO can only use abelian groups, and the $C_{2v}$ subgroup of $C_{\infty v}$ is used for the diatomic molecules. In this group, the $\Sigma^+$ and $\Delta$ states are calculated within the same CI matrix diagonalization and it is sometimes difficult to separate the states of those symmetries. The same is true for the $\Phi$ states, which correspond to the same irreducible representations as the $\Pi$ states in $C_{2v}$.

\section{Non-adiabatic corrections}

We write the total hamiltonian as the sum of an electronic part, $H^{\mathrm{el}}$, and a nuclear kinetic part, $T^{\mathrm{N}}$ which itself can be developed into a radial ($H^{\mathrm{rad}}$) and a rotational ($H^{\mathrm{rot}}$) part:
\begin{eqnarray*}
H & =T^{\mathrm{N}}+H^{\mathrm{el}} \\
	& = H^{\mathrm{rad}} + H^{\mathrm{rot}} + H^{\mathrm{el}}
\end{eqnarray*}
In the electronic hamiltonian, the mass polarization term has been neglected.
Using the electronic wavefunctions $\zeta_{i,\Lambda}$, solutions of the electronic Schr\"odinger equation

\begin{equation*}
H^{\mathrm{el}}\zeta_{i,\Lambda}({\bf r}; R)=U_{i}(R)\zeta_{i,\Lambda}({\bf r}; R) \ ,
\end{equation*} the total wavefunction is expressed as a product of an electronic and a nuclear wavefunction: 
$\Psi({\bf R},{\bf r}) = \sum_{i,\Lambda}\zeta_{i,\Lambda}({\bf r}; R)\psi_{i,\Lambda}({\bf R})$, 
where ${\bf r}$ and ${\bf R}$ stands for the electron and nuclear coordinates, respectively. 
$\Lambda$ is the quantum number associated to $L_z$, the projection of the electronic orbital angular momentum ${\bf L}$ onto the $z$ axis.

As the nuclear hamiltonian is separable, the nuclear wavefunction is given by the product $\psi_{i}({\bf R}) = \psi_{i}(R)\ \vert K\Lambda\rangle$, where $\vert K\Lambda\rangle$ is an eigenfunction of the operators $\mathbf{K}^2$ and $K_{z}$, ${\bf K}$ being the total angular momentum.

Using this development, the Schr\"odinger equation can be expressed as
\begin{equation*}
\sum_{j,\Lambda^{\prime}} \langle\zeta_{i,\Lambda}\vert H^{\mathrm{rad}} + H^{\mathrm{rot}}  \vert\zeta_{j,\Lambda^{\prime}}\rangle 
\psi_{j,\Lambda^{\prime}} + (U_{i}-E)\psi_{i,\Lambda}=0 \ ,
\end{equation*}
where
\begin{equation*}
H^{\mathrm{rad}}=-\frac{1}{2\mu}\partial_{R}^2
\end{equation*}
and
\begin{equation}\label{Hrot}
H^{\mathrm{rot}} = \frac{1}{2\mu R^2}{\bf N}^2 = \frac{1}{2\mu R^2}\Big[{\bf K}^2 + {\bf L}^2 -2K_{z}L_{z} - K_{+}L_{-} - K_{-}L_{+}\Big]
\end{equation}
where ${\bf N}$ is the nuclear angular momentum.

\subsection{Radial couplings}\label{radcoupl}

Using the orthonormality of the electronic wavefunctions, the matrix elements of the radial hamiltonian are given by
\begin{eqnarray*}
\langle\zeta_{i,\Lambda}\vert -\frac{1}{2\mu}\partial_{R}^2 \vert\zeta_{j,\Lambda^{\prime}}\rangle & 
= -\frac{1}{2\mu} \Big[ \partial_{R}^2\delta_{ji} + 2\langle \zeta_{i,\Lambda}\vert \partial_{R}\vert 
\zeta_{j,\Lambda^{\prime}}\rangle\partial_{R} + \langle\zeta_{i,\Lambda}\vert \partial_{R}^2 \vert \zeta_{j,\Lambda^{\prime}}\rangle  
\Big]\delta_{\Lambda^{\prime}\Lambda} \\
	& =-\frac{1}{2\mu}\Big[ \partial_{R}^2\delta_{ji} + 2F_{i\Lambda, j\Lambda^{\prime}}\partial_{R} + 
	G_{i\Lambda, j\Lambda^{\prime}}  \Big]\delta_{\Lambda^{\prime}\Lambda}
\end{eqnarray*}
Since it can be shown that in matrix form $\mathbb{G}=\mathbb{F}^2 + \partial_{R}\mathbb{F}$ (Baer 2006), we only need to calculate the elements of $\mathbb{F}$, which is block-diagonal in $\Lambda$.
These couplings were calculated using a three points central difference method implemented in the DDR program of MOLPRO with a displacement parameter $dR=0.01$ a.u.

\begin{figure}[h!]
\centering
\includegraphics[angle=-90,width=12cm]{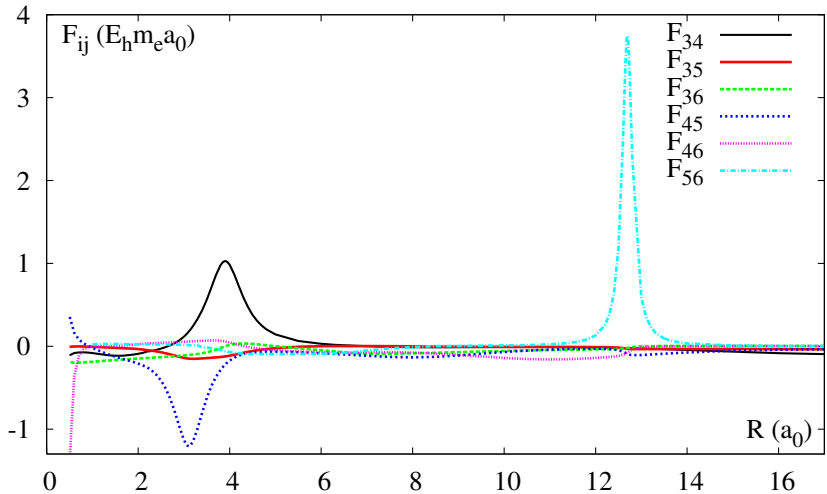}
\caption{Radial non-adiabatic coupling matrix elements between the $n=2$ $^1\Sigma^+$ states. The numbers in subscript refer to the value of $m$ as defined in table \ref{TableSingSig}.}
\label{nacmen=2}
\end{figure}

\begin{figure}[h!]
\centering
\includegraphics[angle=-90,width=12cm]{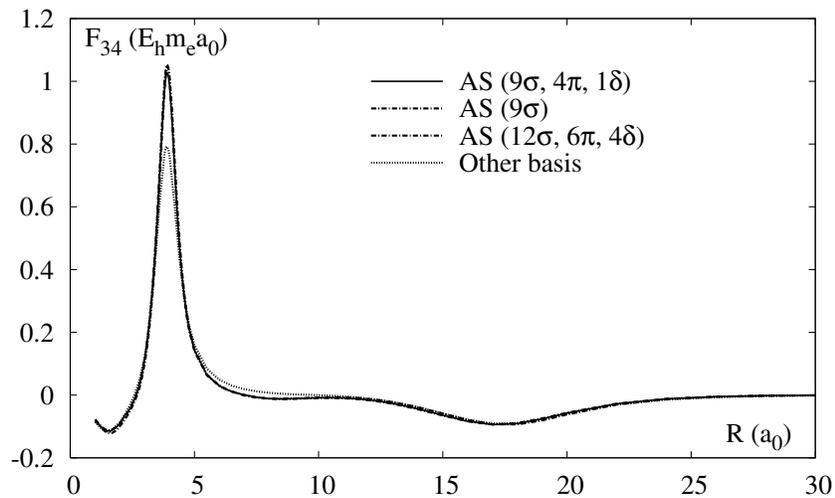}
\caption{Radial non-adiabatic coupling $F_{34}$ calculated with different active spaces (AS), as well as with the basis from van Hemert and Peyerimhoff (1990).}
\label{bosse}
\end{figure}

For the analysis of the radial couplings, we will focus on the $n=2$ states since all the $k(k-1)/2$ couplings (where $k$ is the number of states for a given $\Lambda$) cannot be shown here. 
We notice that the dominant couplings are systematically those connecting two adjacent states ({\it i.e.} the couplings $F_{i,i\pm1}$), as shown in figure \ref{nacmen=2} for the $n=2$ states. This implies that states of different values of $n$ can interact at the exception of the two $n=1$ states which are isolated in energy.
The dominant couplings are narrow and their maxima correspond to the positions of the avoided crossings given in Table \ref{TabNacmeSig} for the $\Sigma$ states and in Table \ref{TabNacmePi} for the $\Pi$ states. The PEC will cross at those points upon diabatization. These couplings are known as ``Landau-Zener couplings'' (Zener 1932). The radial couplings presented in figure \ref{nacmen=2} are of this type. On other hand, some couplings are wider and do not correspond to clear avoided crossings; instead, the PEC are parallel in the coupling region. These couplings arise mainly at large internuclear distances. The dynamics around these couplings is described by the Rosen-Zener theory (Rosen and Zener 1932). Both types of couplings will give rise to very different dynamical behaviors. As an example, we will consider the case of the coupling $F_{34}$ (shown in figure \ref{bosse}), which can be separated in a Landau-Zener coupling (centered around $R\sim 4$ a.u.) and a Rosen-Zener coupling, centered at $R\sim 17$ a.u. The position and the shape of the long-range coupling are invariant under changes in the level of electron correlation or in the atomic basis set used in the calculation, as illustrated in figure \ref{bosse}.

We observed that some of the radial couplings tend asymptotically to a constant which differs from zero. This behaviour is expected for the couplings between two molecular states of same symmetry degenerated at infinity when the calculations of the couplings is done using an origin of the electronic coordinates at the center of the nuclear mass (Belyaev 2001).

\begin{figure}[h!]
\centering
\includegraphics[angle=-90,width=12cm]{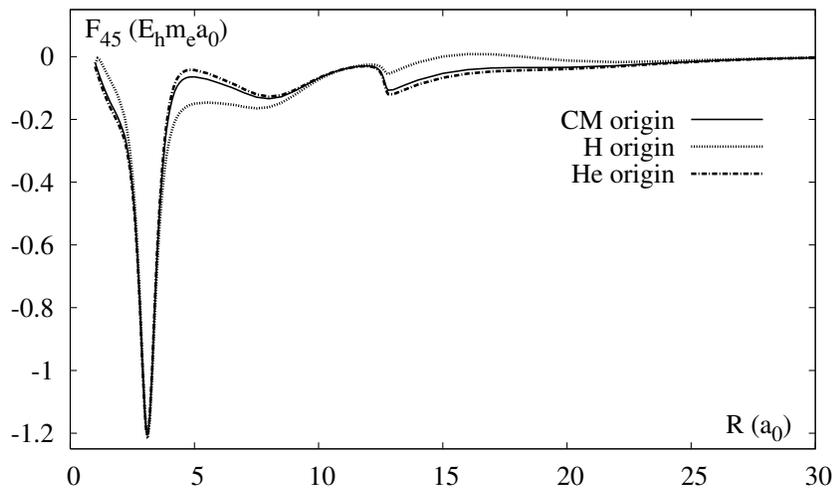}
\caption{Radial non-adiabatic coupling $F_{45}$ calculated using an origin of the coordinates at the center of mass, on H or on He.}
\label{ETF}
\end{figure}

As the final goal of this work is the study of non-adiabatic dynamics involving those couplings, we should note here that a number of authors demonstrated the importance of electron translation factors at high and intermediate energies and proposed different methods to take them into account (Thorson and Delos 1978, Errea \etal 1994). It was also established that these factors are linked to the choice of the origin of the electronic coordinates (Bransden and McDowell 1992). As we noticed the quasi invariance of the radial couplings under a translation of the origin of the electronic coordinates along the internuclear axis (see figure \ref{ETF}), it appears that the inclusion of translation factors in dynamical simulations will not be necessary.

\begin{table}
\caption{\label{TabNacmeSig}Location of the principal non-adiabatic radial couplings for the $n=1-3$ $\Sigma^+$ states.}
\begin{indented}
\item[]
\begin{tabular}{@{}cccccc}
\br
$^1\Sigma^+$ states & $R (a_{0})$ & $^3\Sigma^+$ states & $R (a_{0})$ \\
\mr
	3-4 & 3.9					& 2-3 & 3.6 \\
	4-5 & 3.11 				& 3-4 & 2.9 \\
	5-6 & 12.7 				& 4-5 & 10.6 \\
	6-7 & 7.0					& 5-6 & 3.5, 6.8 \\
	7-8 & 2.9, 5.2	 			& 6-7 & 5.5 \\ 
	8-9 & 4.1					& 7-8 & 4.5 \\
	9-10 & 12.0 				& 8-9 & 10.8, 12.3, 28.0 \\
	9-11 & 12.2 				& 9-10 & 3.6, 12.1, 26.0 \\
	10-11& 4.0, 6.6, 12.2 		& 10-11 & 5.7, 8.8 \\
	11-12 & 2.7, 5.4, 8.8, 25.0 	& & \\
\br
\end{tabular}
\end{indented}
\end{table}

\begin{table}
\caption{\label{TabNacmePi}Location of the principal non-adiabatic radial couplings for the $n=1-3$ $\Pi$ states.}
\begin{indented}
\item[]
\begin{tabular}{@{}cccccc}
\br
$^1\Pi$ states & $R (a_{0})$ & $^3\Pi$ states & $R (a_{0})$ \\
\mr
	3-4 & 7.5			& 3-4 & 7.6  \\
	4-5 & 3.5, 18.8		& 4-5 & 4.5, 18.8 \\
	5-6 & 18.1		& 5-6 & 18.3 \\
\br
\end{tabular}
\end{indented}
\end{table}

\subsection{Rotational coupling}\label{rotcoupl}

From equation (\ref{Hrot}), we can obtain the matrix elements of the rotational hamiltonian between the electronic and rotational nuclear functions. They are given by
\begin{eqnarray}\label{Hrot2} \nonumber
H_{i\Lambda K, j\Lambda^{\prime} K^{\prime}} & = \langle K\Lambda\vert \langle\zeta_{i,\Lambda}\vert H^{\mathrm{rot}}  \vert\zeta_{j,\Lambda^{\prime}}\rangle \vert K^{\prime}\Lambda^{\prime}\rangle \\ \nonumber
& = \frac{1}{2\mu R^2} \bigg\{ \Big[ (K(K+1) - \Lambda^2)\delta_{ij} + \langle\zeta_{i,\Lambda}\vert L_{x}^2+L_{y}^2\vert\zeta_{j,\Lambda^{\prime}}\rangle \Big]\delta_{\Lambda\Lambda^{\prime}} \\
& \qquad +2 \big[ K(K+1)-\Lambda(\Lambda- 1)\big]^{1/2} \langle\zeta_{i,\Lambda}\vert iL_{y}\vert\zeta_{j,\Lambda^{\prime}}\rangle  
\delta_{\Lambda^{\prime},\Lambda+1} \\
& \qquad -2 \big[ K(K+1)-\Lambda(\Lambda + 1) \big]^{1/2} \langle\zeta_{i,\Lambda}\vert iL_{y}\vert\zeta_{j,\Lambda^{\prime}}\rangle 
\delta_{\Lambda^{\prime},\Lambda-1} \bigg\} \delta_{KK^{\prime}} \nonumber
\end{eqnarray}

We see from the formula above that $L_{x}^2+L_{y}^2$ is an interaction between states of the same $\Lambda$ value. In particular, the diagonal part $(L_{x}^2+L_{y}^2)_{ii}$ modifies the energies of the states. This contribution can be evaluated using MOLPRO at the CASSCF level, but we will not report it here since it was shown by Bishop and Cheung (1979) that for the ground state it is of the same order of magnitude as the effect of the mass polarization term, which we have neglected. The same conclusion was reached by Bunker (1968) for H$_2$. We only mention the fact that asymptotically, these matrix elements behave as $R^2$, so that the correction to the energies are constants (see equation (\ref{Hrot2})) when the internuclear distance is large.

The operator $iL_{y}$, on the other hand, connects states with $\Delta\Lambda=\pm 1$ and cannot be neglected. Note that the calculation of the matrix elements of $iL_{y}$ necessitates the simultaneous determination of electronic states of different value of $\Lambda$ in the same calculation.

\begin{figure}[h!]
\centering
\includegraphics[angle=-90,width=16cm]{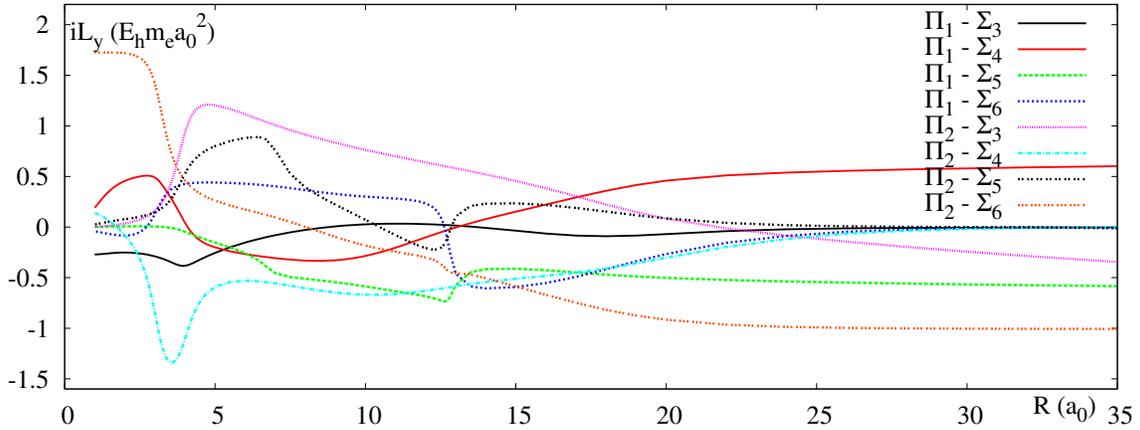}
\caption{Adiabatic rotational couplings between the $n=2$  $^1\Sigma^+$ and $^1\Pi$ states. The numbers in subscript refer to the value of $m$ as defined in table \ref{TableSingSig}.}
\label{rotn=2}
\end{figure}

The rotational couplings between the $n=2$ $^1\Pi$ and $^1\Sigma^+$ states are presented in figure \ref{rotn=2}. Asymptotically, the couplings between states that dissociate into the same atomic species and into the same $n$ manifold are constant. Some rotational couplings between states dissociating into different $n$ manifolds make an exception and behave asymptotically as $R$, as indicated by Belyaev \etal (2001). All the couplings between states dissociating into different atomic species tend to zero at large internuclear distances.

\section{Adiabatic dipole moments}

\begin{figure}[h!]
\centering
\includegraphics[angle=-90,width=13cm]{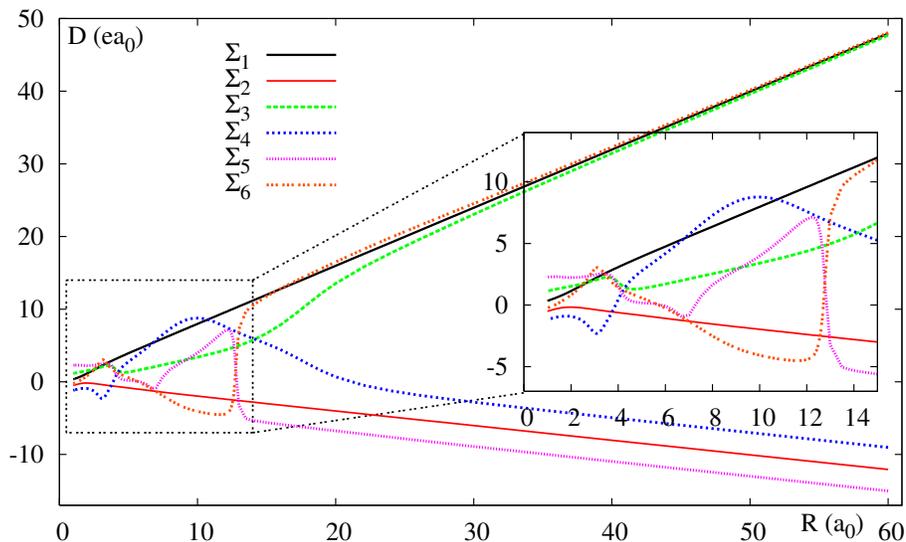}
\caption{Permanent dipole moments for the $n=1,2$ $^1\Sigma^+$ states.}
\label{permdip}
\end{figure}

\begin{figure}[h!]
\centering
\includegraphics[angle=-90,width=13cm]{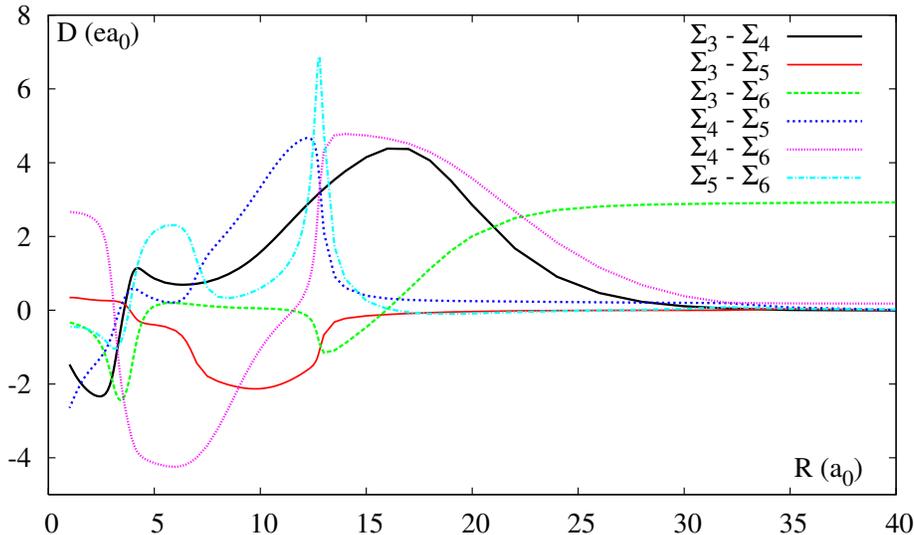}
\caption{Transition dipole moments between the $n=2$ $^1\Sigma^+$ states.}
\label{transdip}
\end{figure}

There are $k(k+1)/2$ (where $k$ is the number of states) dipole matrix elements and we will again restrict our discussion to the $^1\Sigma^+$ states. The permanent adiabatic dipole moments of the $n=1,2$ states are represented in figure \ref{permdip}, while the transition dipole moments of the $n=2$ states are represented in figure \ref{transdip}. 

In a given $\Lambda$ subspace, the dipole interaction occurs only through the $z$ component. As there is no ambiguity, we will thus write the dipole moment between two states $i$ and $j$ as $\mu_{ij}$ instead of $\mu_{z, ij}$. 

The behavior of the dipole moments is consistent with the calculation of the radial couplings, illustrating the relation between the two operators (Macias and Riera 1978) which allows the use of the dipole moment rather than the radial couplings to find a diabatic representation. For example, the crossing between the dipoles $\mu_{55}$ and $\mu_{66}$ at $R=12.75$ a.u. correspond to the sharp radial coupling seen in figure \ref{nacmen=2} at the same internuclear distance, and is reflected on the transition dipole moment between the two states, $\mu_{56}$, which presents a sharp peak in the crossing region. Conversely, when two permanent dipole have a peak of opposite value, as do $\mu_{44}$ and $\mu_{55}$ at $R=3.1$ a.u., the sign of the transition dipole $\mu_{45} $ changes abruptly. Again, this is linked to an avoided crossing between the fourth and fifth states at the same internuclear distance.

As can be seen from figure \ref{permdip}, all the permanent dipole moments $\mu_{ii}$ behave asymptotically as $R$, since the origin of the coordinates is at the nuclear center of mass and not on one of the atoms. As the reduced mass of $^4$HeH$^+$ is $\mu=0.805$ a.m.u., the helium and the hydrogen nuclei are situated approximatively at $-0.2R$ and $0.8R$ of the origin, respectively. The permanent dipole moment can then be divided into a nuclear and an electronic part. The nuclear contribution is of $0.4R$ and is identical for all states, while the electronic contribution is $0.4R$ for the He($1snl\ ^1L$) + H$^+$ states and of $-0.6R$ for the He$^+(1s)$ + H$(nl)$ states. For the latter, there is an additional contribution from the interaction between the helium $1s$ electron and the hydrogen $nl$ electron which explains that the permanent dipole moments for these states do not tend to the same asymptotic value.

\section{Diabatic representation of the Rydberg states}

The diabatic representation is defined so as to cancel the $\mathbb{F}$ matrix, which is the case if the adiabatic-to-diabatic transformation matrix $\mathbb{D}$ satisfies the matrix equation 
\begin{equation}\label{ADT}
\partial_R\mathbb{D}+\mathbb{F\cdot D}=0
\end{equation} 
The diabatic potential energy curves are then given as the diagonal elements of the transformed matrix $\mathbb{U}^\mathrm{d}=\mathbb{D}^{-1}\cdot \mathbb{U\cdot D}$. We solve equation (\ref{ADT}) by continuity using a grid of 2000 points, starting from $R=60$ a.u. where we require that the adiabatic and diabatic representations are identical (so that $\mathbb{D=I}$). It should also be noted that, as we calculate the non-adiabatic radial coupling at the CASSCF level, we also use the CASSCF energies, which differ slightly from the CI energies presented in section \ref{PEC}.

In our diabatization procedure, we will not consider the complete $\mathbb{F}$ matrix, keeping only the couplings $F_{i,i+1}$ and putting all the other couplings to zero. This approximation, which amounts to a succession of two-states cases, is used for various reasons. 
The first one is that, as was shown in section \ref{radcoupl}, those couplings are systematically the most important ones. 
Secondly, it has been shown by Zhu and Nakamura (1997) that this approximation gives correct results in dynamical calculations even for low energies, which is confirmed by our calculations of the electron-transfer cross sections in the $n=2$ manifold (Loreau \etal 2010). 
Thirdly, some of the couplings which we neglect remain non-zero at large internuclear distance. This phenomenon is known and has been discussed in detail by Belyaev \etal (2001), but raises a problem in our diabatization method. Indeed, our procedure is based on the fact that the adiabatic and diabatic representation coincide at $R=\infty$, which is not the case if the couplings do not vanish. As a consequence, these residual couplings, although small, influence the diabatic PEC at large $R$, a feature which is of course undesirable.

\begin{figure}[h!]
\centering
\includegraphics[angle=-90,width=14cm]{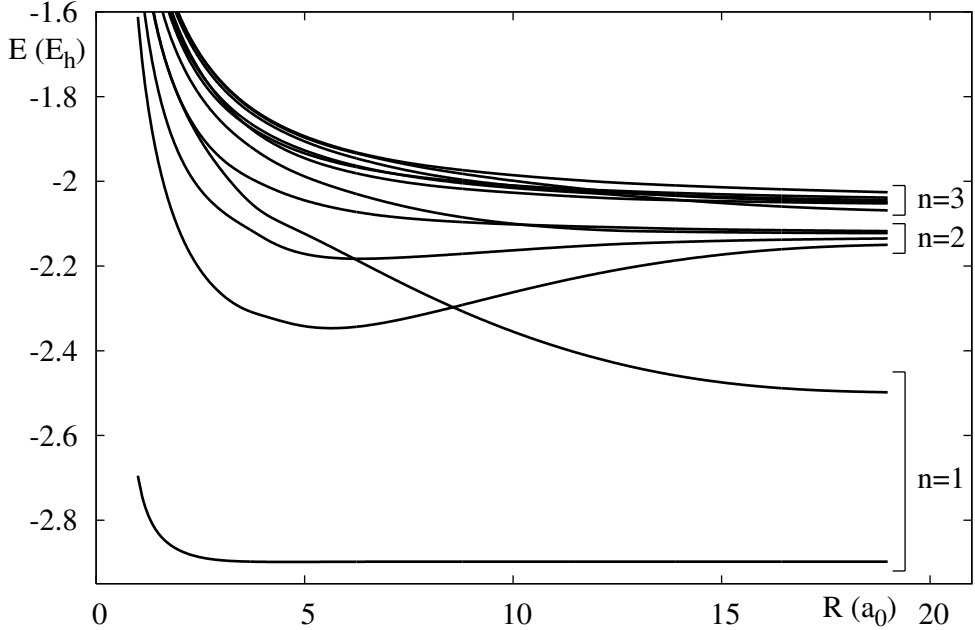}
\caption{Diabatic PEC of the $n=1-3$ $^1\Sigma^+$ states.}
\label{diabn=1+2+3}
\end{figure}

We will consider the $^1\Sigma^+$ states as an example of the diabatization procedure. The result of the diabatization for these states up to $n=3$ is given as an example in figure \ref{diabn=1+2+3}. 

\begin{figure}[h!]
\centering
\includegraphics[angle=-90,width=12cm]{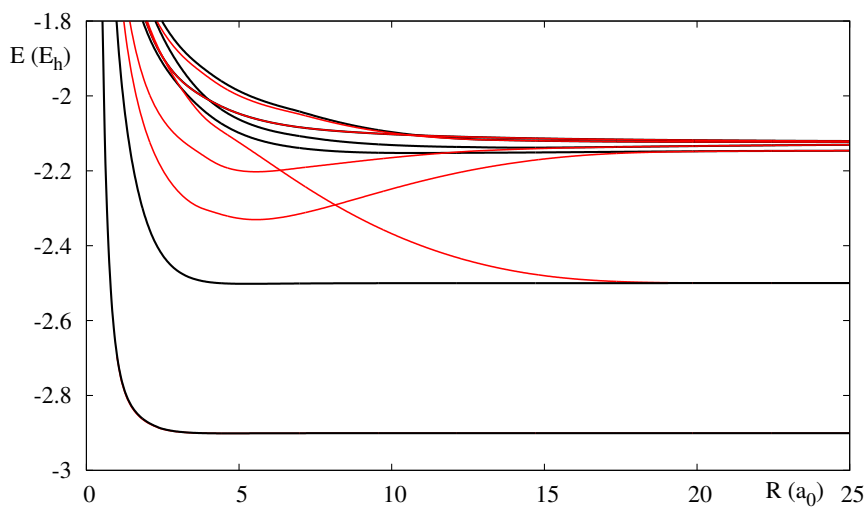}
\caption{Comparison of the diabatic PEC of the $n=1$ and $n=2$ $^1\Sigma^+$ states diabatized independently  (black lines) or as a whole (red lines).}
\label{diabn=1+2}
\end{figure}

It is clear that the diabatization alters considerably the shape of the ground state. From figure \ref{diabn=1+2}, we see that this is essentially due to the coupling $F_{12}$ since the same behaviour is observed when the $n=1$ states are diabatized independently. The first excited state also changes dramatically, crossing the entire $n=2$ manifold in the diabatic representation, which is due to the coupling $F_{23}$ (see figure \ref{diabn=1+2}). However, one must remember that the non-adiabatic coupling between the two first $^1\Sigma^+$ states is of the Rosen-Zener type, and that therefore the diabatic representation of those states has little physical significance. Likewise, the influence of the two first $^1\Sigma^+$ states on the diabatic representation of the $n=2$ states is very important but has little effect on the non-adiabatic dynamics in the $n=2$ manifold, as has been observed in the calculation of charge exchange cross section between He$^+$ and H at low energies (Loreau \etal 2010). To our knowledge, the only other work on the diabatic representation of the PEC of the HeH$^+$ ion has been done by Kl¨\"uner \etal (1999) using the quasi-diabatization procedure proposed by Pacher \etal (1988). These authors compare a 3-states (the three lowest $n=2$ states) and a 4-states (adding the second $n=1$ state) diabatization in a small interval of internuclear distances (0.8 a.u. $\leq R \leq$ 5.4 a.u.). It is concluded that the inclusion of the $n=1$ state does not modify the diabatic PEC of the $n=2$ manifold, and this state is therefore neglected in wavepacket simulations of charge exchange processes involving $n=2$ states. Although we arrive at the same conclusion regarding the dynamics, our method gives significantly different diagonal as well as non-diagonal matrix elements of the electronic hamiltonian in the diabatic representation.

\begin{figure}[h!]
\centering
\includegraphics[angle=-90,width=12cm]{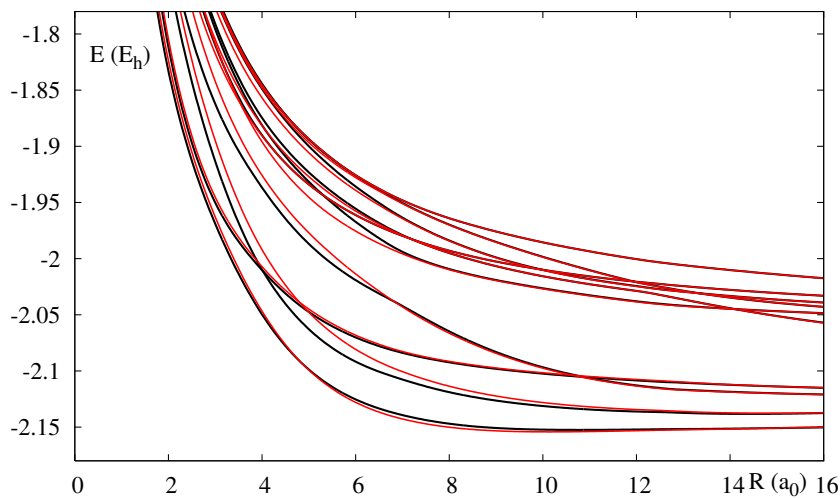}
\caption{Comparison of the diabatic PEC of the $n=2$ and $n=3$ $^1\Sigma^+$ states diabatized independently  (black lines) or as a whole (red lines).}
\label{diabn=2+3}
\end{figure}

On the other hand, the effect of the interaction between the $n=2$ and $n=3$ manifolds through the $F_{67}$ matrix element is relatively small, as shown in figure \ref{diabn=2+3}. 

\begin{figure}[h!]
\centering
\includegraphics[angle=-90,width=12cm]{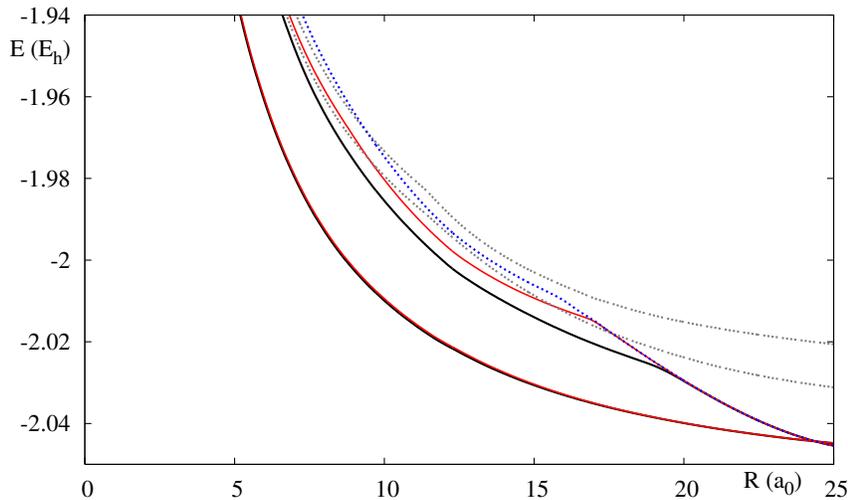}
\caption{Effect of the $n=4$ states on the six $n=3$ states. In black, the fifth and sixth $n=3$ diabatic states. In red, the same states, but diabatized with the first $n=4$ state. The fifth $n=3$ state is left unaltered, but the avoided crossing of the sixth $n=3$ state is shifted from $R=20$ au to $R=17$ au. In blue, the sixth $n=3$ state diabatized with the first and second $n=4$ state (dashed grey line). The avoided crossing is further shifted to $R=16$ au.}
\label{diab14}
\end{figure}

Finally, the description of the diabatic $n=3$ states necessitates to take some higher-lying states into account, since the highest diabatic state undergoes an avoided crossing around $R=20$ (as can be seen in figure \ref{pot1}) with the first $n=4$ state. The inclusion of the first two $n=4$ states in the diabatization, while leaving the first five $n=3$ states unaltered, clearly influence the sixth state by shifting the position of the avoided crossing to smaller values of $R$, as illustrated in figure \ref{diab14}. Therefore, a more correct description of the last $n=3$ diabatic state should include more $n=4$ states, even though we are only considering the couplings between adjacent states.

\section{Conclusions}


We present an accurate description of 66 low lying adiabatic states of HeH$^+$. Using the MOLPRO package and a large adapted basis set, the potential energy curves of the $n=1-3$ as well as most of the $n=4$ states of the molecular ion have been obtained and compared to previous theoretical works. The radial and rotational non-adiabatic coupling matrix elements, as well as the dipole matrix elements, have been calculated for all the $n=1-3$ states. The radial couplings allow to switch to the diabatic representation which is used to treat dynamical processes involving the ion. This material has been used to calculate the cross section of the photodissociation of the ion (Sodoga \etal 2009) and the cross sections for the charge transfer processes He$^+(1s)$ + H($nl) \longrightarrow $ He($1sn^{\prime}l^{\prime}\ ^{1,3}L)$ + H$^+$ at low energy (Loreau \etal 2010).

\ack

We thank M. Godefroid for helpful discussions. This work was supported by the Fonds National de la Recherche Scientifique (IISN projects) and by the “Action de Recherche Concert\'ee” ATMOS de la Communaut\'e Fran\c caise de Belgique. P. P. and P. Q. are respectively research associate and senior research associate of the Belgian F.R.S.-F.N.R.S. J. L. thanks the FRIA for financial support.

\newpage
\appendix
\section{Basis set}

\begin{table}[h]
\caption{\label{Basis_H} Additional basis functions for hydrogen.}
\begin{indented}
\item[]
\begin{tabular}{@{}ccccccc}
\br
	& Coefficient	& Exponent	& & 	& Coefficient	& Exponent	\\
\mr
$2s$	& 0.000144 & 19.907407	& & $3p$	& 0.024590 & 0.986107	\\
	& -0.005010 & 8.988620 	& &		& 0.260387 & 0.181062 	\\
	& -0.059868 & 0.645915	& & 		& 0.903505 & 0.051409 	\\
	& -0.123387 & 0.071221	& & 		& -1.874014 & 0.007783 	\\
	& 0.891720 & 0.024689	& & 		& -0.287389 & 0.003281 	\\
	&		&			& & 		& & \\
$3s$	& 0.017083 & 1.000840	& & $4p$	& 0.015689 & 1.015267	\\
	& 0.120859 & 0.272819	& & 		& 0.170182 & 0.186800	\\
	& 0.381641 & 0.093336	& & 		& 0.625077 & 0.053622	\\
	& 0.782727 & 0.023846	& & 		& -2.040089 & 0.009383	\\
	& -2.036422 & 0.019287 	& & 		& 2.535609 & 0.002477	\\
	& 1.594513 & 0.004250	& & 		& & \\
	&		&			& & 	$3d$	& 0.050502 & 0.199534	\\	
$4s$	& 0.013653 & 11.939994	& & 		& 0.495617 & 0.051732	\\
	& 0.167164 & 1.060859	& & 		& 1.609201 & 0.018241	\\
	& 0.662095 & 0.188361	& & 		& 1.213582 & 0.007264	\\
	& -3.032473 & 0.029081	& & 		& & 	\\
	& 4.828012 & 0.007375	& & $4d$	& 0.077845 & 0.083401	\\
	& -4.362099 & 0.001197	& & 		& 0.454482 & 0.021517	\\
	&		&			& & 		& -0.894944 & 0.003285	\\
$2p$	& 0.005698 & 3.101143	& & 		& & 	\\
	& 0.067894 & 0.567270	& & $4f$	& 0.026401 & 0.033813	\\
	& 0.407521 & 0.158596	& & 		& 0.169866 & 0.010591	\\
	& 1.082414 & 0.052537	& & 		& 0.204085 & 0.004109	\\
	& 0.800233 & 0.02001	& & 		& &	\\
\br
\end{tabular}
\end{indented}
\end{table}

\begin{table}[h]
\caption{\label{Basis_H} Additional basis functions for the singlet states of helium.}
\begin{indented}
\item[]
\begin{tabular}{@{}ccccccc}
\br
	& Coefficient	& Exponent	& & 	& Coefficient	& Exponent	\\
\mr
$2s$	& 0.002760 & 99.181545	& & $3p$	& 0.002929 & 4.673075	\\
	& 0.030509 & 10.536516 	& &		& 0.037026 & 0.848515 	\\
	& 0.196191 & 1.905592	& & 		& 0.244477 & 0.227656 	\\
	& 0.775469 & 0.424822	& & 		& 0.806379 & 0.070757 	\\
	& -3.722470 & 0.029920	& & 		& -2.050784 & 0.007911 	\\
	&		&			& & 		& & \\
$3s$	& -0.001952 & 99.838262	& & $4p$	& 0.002547 & 4.629236	\\
	& 0.012286 & 49.068205	& & 		& 0.076795 & 0.396874	\\
	& 0.151694 & 2.653277	& & 		& 0.488599 & 0.081760	\\
	& 0.516699 & 0.309173	& & 		& 3.304293 & 0.017020	\\
	& -3.156883 & 0.038398	& & 		& -4.848362 & 0.013667	\\
	& 4.383564 & 0.006064	& & 		& 2.305369 & 0.002285	\\
	&		&			& & 		& &	\\	
$4s$	& -0.004235 & 99.861298	& & $3d$	& 0.062634 & 0.102716	\\
	& 0.011613 & 49.188130	& & 		& 0.430990 & 0.026362	\\
	& 0.316811 & 0.732304	& & 		& 0.551594 & 0.008950	\\
	& -2.495917 & 0.035479	& & 		& & 	\\
	& 4.534324 & 0.009221	& & $4d$	& 0.077839 & 0.082347	\\
	& 4.267928 & 0.001472	& & 		& 0.454447 & 0.021503	\\
	&		&			& & 		& -0.894818 & 0.003283	\\
$2p$	& 0.000152 & 19.984152	& & 		& & 	\\
	& 0.002135 & 4.719990	& & $4f$	& 0.026447 & 0.033780	\\
	& 0.025319 & 1.021806	& & 		& 0.170021 & 0.010582	\\
	& 0.160069 & 0.292204	& & 		& 0.204038 & 0.004106	\\
	& 0.592704 & 0.101345	& & 		& &	\\
	& 1.066547 & 0.039590	& & 		& & 	\\
	& 0.527114 & 0.016351	& & 		& & 	\\
\br
\end{tabular}
\end{indented}
\end{table}

\begin{table}
\caption{\label{Basis_H} Additional basis functions for the triplet states of helium.}
\begin{indented}
\item[]
\begin{tabular}{@{}ccccccc}
\br
	& Coefficient	& Exponent	& & 	& Coefficient	& Exponent	\\
\mr
$2s$	& 0.006837 & 99.181525	& & $3p$	& 0.002497 & 4.675067	\\
	& 0.067529 & 10.551100 	& &		& 0.035791 & 0.789129 	\\
	& 0.269982 & 2.355836	& & 		& 0.247261 & 0.205234 	\\
	& 0.759139 & 0.643556	& & 		& 0.833266 & 0.065732 	\\
	& -3.666553 & 0.037246	& & 		& -2.061784 & 0.007542	\\
	&		&			& & 		& & \\
$3s$	& -0.000849 & 99.836273	& & $4p$	& 0.013260 & 1.567909	\\
	& 0.012336 & 49.044874	& & 		& 0.137119 & 0.284565	\\
	& 0.156413 & 3.473103	& & 		& 0.208845 & 0.076404	\\
	& 0.448487 & 0.589924	& & 		& 3.155372 & 0.015570	\\
	& -3.036160 & 0.043579	& & 		& -4.866015 & 0.013266	\\
	& 4.329233 & 0.006971	& & 		& 2.376480 & 0.002506	\\
	&		&			& & 		& &	\\	
$4s$	& -0.004869 & 99.860674	& & $3d$	& 0.016008 & 0.199406	\\
	& 0.015055 & 49.185022	& & 		& 0.156978 & 0.051714	\\
	& 0.327369 & 1.070000	& & 		& 0.508950 & 0.018239	\\
	& -2.490977 & 0.042209	& & 		& 0.383308 & 0.007266 	\\
	& 4.623855 & 0.011100	& & 		& &	\\
	& -4.295313 & 0.001760	& & $4d$	& 0.010397 & 0.218719	\\
	&		&			& & 		& 0.112909 & 0.056369	\\
$2p$	& 0.005913 & 4.650435	& & 		& 0.423743 & 0.019585 	\\
	& 0.158917 & 0.458934	& &		& -0.906780 & 0.003352	\\
	& 0.939676 & 0.102736	& & 		& &	\\
	& 1.256290 & 0.030224	& & $4f$	& 0.018968 & 0.038973	\\
	&		&			& & 		& 0.139803 & 0.012212	\\	
	&		&			& & 		& 0.230793 & 0.004712	\\
	&		&			& & 		& 0.999016 & 0.000133	\\	
\br
\end{tabular}
\end{indented}
\end{table}

\newpage

\section*{References}
\begin{harvard}
\item[] Badnell N R 1986 \JPB {\bf 9} 3827
\item[] Badnell N R 1997 \jpb {\bf 30} 1
\item[] Baer M 2006 {\it Beyond Born-Oppenheimer} (John Wiley \& Sons) p 9
\item[] Belyaev AK, Egorova D, Grosser J and Menzel T 2001 {\it Phys. Rev. A \bf 64} 052701
\item[] Bishop DM and Cheung LM 1979 {\it J. Mol. Spec. \bf 75} 462
\item[] Boutalib A and Gadea FX 1992 {\it J. Chem. Phys. \bf 97} 1144
\item[] Bransden BH and McDowell MR 1992 {\it Charge Exchange and the Theory of Ion-Atom Collisions} (Oxford: Clarendon Press)
\item[] Bunker PR 1968 {\it J. Mol. Spectrosc. \bf 28} 422
\item[] Chibisov MI, Yousif FB, Van der Donk PJT and Mitchell JBA 1996 {\it Phys. Rev. A \bf 54} 5997
\item[] Coxon JA and Hajigeorgiou G 1999 {\it J. Mol. Spec. \bf 193} 306
\item[] Dunning TH 1989 {\it J. Chem. Phys. \bf 90} 1007
\item[] Eissner W, Jones M and Nussbaumer H 1974 {\it Comput. Phys. Commun. \bf 8} 270
\item[] Engel EA, Doss N, Harris GJ and Tennyson J, 2005 {\it Mon. Not. R. Astron. Soc.} {\bf 357} 471
\item[] Errea LF, Harel C, Jouin H, M\'endez L, Pons B and Riera A 1994 \jpb {\bf 27} 3603
\item[] Frebel A \etal 2005 {\it Nature} {\bf 434} 871
\item[] Giesbertz KJH, Baerends EJ and Gritsenko OV 2008 {\it Phys. Rev. Lett.} {\bf 101} 033004
\item Goldman and Cassar 2005 Atoms in Strong Fields {\it Handbook of Atomic, Molecular and Optical Physics}, Drake (ed), (New York: Springer)
\item[] Green TA,  Browne JC, Michels HH and Madsen MM 1974a {\it J. Chem. Phys. \bf 61} 5186
\item[] Green TA,  Browne JC, Michels HH and Madsen MM 1974b {\it J. Chem. Phys. \bf 61} 5198
\item[] Green TA, Michels HH and Browne JC 1976 {\it J. Chem. Phys. \bf 64} 3951
\item[] Green TA, Michels HH and Browne JC 1978 {\it J. Chem. Phys. \bf 69} 101
\item[] Harris GJ, Lyans-Gray AE, Miller S and Tennyson J 2004 {\it Ap. J.} {\bf 617} L143
\item[] Hogness TR and Lunn EC 1925 {\it Phys. Rev.} {\bf 26} 44
\item[] Jurek M, Spirko V and Kraemer WP 1995 {\it Chem. Phys. \bf 193} 287
\item[] Ketterle W, Figger H and Walther H 1985 {\it Phys. Rev. Lett.} {\bf 55} 2941
\item[] Kl¨\"uner T, Thiel S and Staemmler V 1999 \jpb {\bf 32} 4931
\item[] Knowles PJ and Werner H-J 1985 {\it Chem. Phys. Lett. \bf 15} 259
\item[] Kolos W 1976 {\it Int. J. Quantum Chem. \bf X} 217
\item[] Kolos W and Peek JM 1976 {\it Chem. Phys. \bf 12} 381
\item[] Larson \AA \ and Orel AE 2005 {\it Phys. Rev. A} {\bf 72} 032701
\item[] Loreau J, Desouter-Lecomte M, Sodoga K, Lauvergnat D and Vaeck N 2010 {\it Phys. Rev. A}, in preparation.
\item[] Lepp S, Stancil PC and Dalgarno A, 2002 \jpb {\bf 35} R57
\item[] Mac\'ias A and Riera A 1978 \jpb {\bf 11} L489
\item[] Michels HH 1966 {\it J. Chem. Phys. \bf 44} 3834
\item[] Mills R and Ray P 2003 {\it J. Phys. D: Appl. Phys.} {\bf 36} 1535
\item[] Moorhead JM, Lowe RP, Maillard J-P, Wehlau WH and Bernath PF 1988 {\it Ap. J.} {\bf 326} 899
\item[] Pacher T, Cederbaum L S and Köppel H 1988 {\it J. Chem. Phys. \bf 89} 7367
\item[] Pedersen HB \etal 2007 {\it Phys. Rev. Lett. \bf 98} 223202
\item Radzig AA and Smirnov BM 1985 {\it Reference Data on Atoms, Molecules and Ion} (Berlin: Springer)
\item[] Richings GW and Karadakov PB 2007 {\it Mol. Phys.} {\bf 105} 2363
\item[] Roberge W and Dalgarno A 1982 {\it Ap. J.} {\bf 255} 489
\item[] Rosen N and Zener C 1932 {\it Phys. Rev.} {\bf 40} 502
\item[] Rosmej, FB, Stamm E and Lisitsa VS 2006 {\it Europhys. Lett.} {\bf 73} 342
\item[] Sodoga K, Loreau J, Lauvergnat D, Justum Y, Vaeck N and Desouter-Lecomte M 2009 {\it Phys. Rev. A \bf 80} 033417
\item[] Stanke M, Kedziera D, Molski M, Bubin S, Barysz M and Adamowicz L 2006 {\it Phys. Rev. Lett. \bf 96} 	233002
\item[] Stanke M, Kedziera D, Bubin S and Adamowicz L 2008 {\it Phys. Rev. A \bf 77} 022506
\item[] Thorson WR and Delos JB 1978 {\it Phys. Rev. A} {\bf 18} 117
\item[] van Hemert MC and Peyerimhoff SD 1990 {\it J. Chem. Phys.} {\bf 94} 4369
\item[] Werner H-J and Knowles PJ 1985 {\it J. Chem. Phys. \bf 82} 5053
\item[] Werner H-J, Knowles PJ, Lindh R, Manby FR, Sch¨\"utz M, and others. MOLPRO, version 2006.1, a package of {\it ab initio} programs, see http://www.molpro.net
\item[] Woon DE  and Dunning TH 1994 {\it J. Chem. Phys. \bf 100} 2975
\item[] Yan ZC 2000 {\it Phys. Rev. A \bf 62} 052502
\item[] Yan ZC 2002\jpb {\bf 35} 2713
\item[] Zener C 1932 {\it Proc. R. Soc. \bf 137} 696
\item[] Zhu C and Nakamura H 1997 {\it J. Chem. Phys. \bf 106} 2599

\end{harvard}

\end{document}